\documentclass[twocolumn]{aastex631}

\usepackage{graphicx}
\usepackage{amsmath}
\usepackage{multirow}
\usepackage{todonotes}
\usepackage{float}

\usepackage{savesym}
\savesymbol{tablenum}
\usepackage{siunitx}
\restoresymbol{SIX}{tablenum}

\DeclareSIUnit\year{yr}
\usepackage{booktabs}

\begin{document}

\title{
Investigation of the Microquasar SS 433 with VERITAS
}

\collaboration{200}{The VERITAS collaboration}

\author{A.~Archer}\affiliation{Department of Physics and Astronomy, DePauw University, Greencastle, IN 46135-0037, USA}
\author[0000-0002-3886-3739]{P.~Bangale}\affiliation{Department of Physics, Temple University, Philadelphia, PA 19122, USA}
\author[0000-0002-9675-7328]{J.~T.~Bartkoske}\affiliation{Department of Physics and Astronomy, University of Utah, Salt Lake City, UT 84112, USA}
\author[0000-0003-2098-170X]{W.~Benbow}\affiliation{Center for Astrophysics $|$ Harvard \& Smithsonian, Cambridge, MA 02138, USA}
\author{N.~R.~Bond}\affiliation{School of Physics, University College Dublin, Belfield, Dublin 4, Ireland}
\author[0009-0001-5719-936X]{Y.~Chen}\affiliation{Department of Physics and Astronomy, University of California, Los Angeles, CA 90095, USA}
\author[0000-0001-5811-9678]{J.~L.~Christiansen}\affiliation{Physics Department, California Polytechnic State University, San Luis Obispo, CA 94307, USA}
\author{A.~J.~Chromey}\affiliation{Center for Astrophysics $|$ Harvard \& Smithsonian, Cambridge, MA 02138, USA}
\author[0000-0003-1716-4119]{A.~Duerr}\affiliation{Department of Physics and Astronomy, University of Utah, Salt Lake City, UT 84112, USA}
\author[0000-0002-1853-863X]{M.~Errando}\affiliation{Department of Physics, Washington University, St. Louis, MO 63130, USA}
\author{M.~E.~Godoy}\affiliation{Santa Cruz Institute for Particle Physics and Department of Physics, University of California, Santa Cruz, CA 95064, USA}
\author[0000-0002-4131-655X]{J.~E.~Pedrosa}\affiliation{Center for Astrophysics $|$ Harvard \& Smithsonian, Cambridge, MA 02138, USA}
\author{S.~Feldman}\affiliation{Department of Physics and Astronomy, University of California, Los Angeles, CA 90095, USA}
\author[0000-0001-6674-4238]{Q.~Feng}\affiliation{Department of Physics and Astronomy, University of Utah, Salt Lake City, UT 84112, USA}
\author[0000-0002-2636-4756]{S.~Filbert}\affiliation{Department of Physics and Astronomy, University of Utah, Salt Lake City, UT 84112, USA}
\author[0000-0003-1614-1273]{A.~Furniss}\affiliation{Santa Cruz Institute for Particle Physics and Department of Physics, University of California, Santa Cruz, CA 95064, USA}
\author[0000-0002-0109-4737]{W.~Hanlon}\affiliation{Center for Astrophysics $|$ Harvard \& Smithsonian, Cambridge, MA 02138, USA}
\author[0000-0003-3878-1677]{O.~Hervet}\affiliation{Santa Cruz Institute for Particle Physics and Department of Physics, University of California, Santa Cruz, CA 95064, USA}
\author[0000-0001-6951-2299]{C.~E.~Hinrichs}\affiliation{Center for Astrophysics $|$ Harvard \& Smithsonian, Cambridge, MA 02138, USA and Department of Physics and Astronomy, Dartmouth College, 6127 Wilder Laboratory, Hanover, NH 03755 USA}
\author[0000-0002-6833-0474]{J.~Holder}\affiliation{Department of Physics and Astronomy and the Bartol Research Institute, University of Delaware, Newark, DE 19716, USA}
\author[0000-0002-1432-7771]{T.~B.~Humensky}\affiliation{Department of Physics, University of Maryland, College Park, MD, USA and NASA GSFC, Greenbelt, MD 20771, USA}
\author{M.~Iskakova}\affiliation{Department of Physics, Washington University, St. Louis, MO 63130, USA}
\author[0000-0002-1089-1754]{W.~Jin}\affiliation{Department of Physics and Astronomy, University of California, Los Angeles, CA 90095, USA}
\author[0009-0008-2688-0815]{M.~N.~Johnson}\affiliation{Santa Cruz Institute for Particle Physics and Department of Physics, University of California, Santa Cruz, CA 95064, USA}
\author[0000-0001-8557-1141]{E.~Joshi}\affiliation{Deutsches Elektronen-Synchrotron DESY, Platanenallee 6, 15738 Zeuthen, Germany}
\author[0000-0002-3638-0637]{P.~Kaaret}\affiliation{Department of Physics and Astronomy, University of Iowa, Van Allen Hall, Iowa City, IA 52242, USA}
\author{M.~Kertzman}\affiliation{Department of Physics and Astronomy, DePauw University, Greencastle, IN 46135-0037, USA}
\author{M.~Kherlakian}\affiliation{Fakult\"at f\"ur Physik \& Astronomie, Ruhr-Universit\"at Bochum, D-44780 Bochum, Germany}
\author[0000-0002-4260-9186]{T.~K.~Kleiner}\affiliation{Deutsches Elektronen-Synchrotron DESY, Platanenallee 6, 15738 Zeuthen, Germany}
\author[0000-0002-4289-7106]{N.~Korzoun}\affiliation{Department of Physics and Astronomy and the Bartol Research Institute, University of Delaware, Newark, DE 19716, USA}
\author{F.~Krennrich}\affiliation{Department of Physics and Astronomy, Iowa State University, Ames, IA 50011, USA}
\author[0000-0002-5167-1221]{S.~Kumar}\affiliation{Department of Physics, University of Maryland, College Park, MD, USA }
\author{S.~Kundu}\affiliation{Department of Physics and Astronomy, University of Alabama, Tuscaloosa, AL 35487, USA}
\author[0000-0003-4641-4201]{M.~J.~Lang}\affiliation{School of Natural Sciences, University of Galway, University Road, Galway, H91 TK33, Ireland}
\author[0000-0003-3802-1619]{M.~Lundy}\affiliation{Physics Department, McGill University, Montreal, QC H3A 2T8, Canada}
\author[0000-0001-9868-4700]{G.~Maier}\affiliation{Deutsches Elektronen-Synchrotron DESY, Platanenallee 6, 15738 Zeuthen, Germany}
\author[0000-0002-1499-2667]{P.~Moriarty}\affiliation{School of Natural Sciences, University of Galway, University Road, Galway, H91 TK33, Ireland}
\author[0000-0002-3223-0754]{R.~Mukherjee}\affiliation{Department of Physics and Astronomy, Barnard College, Columbia University, NY 10027, USA}
\author{M.~Ohishi}\affiliation{Institute for Cosmic Ray Research, University of Tokyo, 5-1-5, Kashiwa-no-ha, Kashiwa, Chiba 277-8582, Japan}
\author[0000-0002-4837-5253]{R.~A.~Ong}\affiliation{Department of Physics and Astronomy, University of California, Los Angeles, CA 90095, USA}
\author[0000-0003-3820-0887]{A.~Pandey}\affiliation{Department of Physics and Astronomy, University of Utah, Salt Lake City, UT 84112, USA}
\author[0000-0001-7861-1707]{M.~Pohl}\affiliation{Institute of Physics and Astronomy, University of Potsdam, 14476 Potsdam-Golm, Germany and Deutsches Elektronen-Synchrotron DESY, Platanenallee 6, 15738 Zeuthen, Germany}
\author[0000-0002-0529-1973]{E.~Pueschel}\affiliation{Fakult\"at f\"ur Physik \& Astronomie, Ruhr-Universit\"at Bochum, D-44780 Bochum, Germany}
\author[0000-0002-5104-5263]{P.~L.~Rabinowitz}\affiliation{Department of Physics, Washington University, St. Louis, MO 63130, USA}
\author[0000-0002-5351-3323]{K.~Ragan}\affiliation{Physics Department, McGill University, Montreal, QC H3A 2T8, Canada}
\author{P.~T.~Reynolds}\affiliation{Department of Physical Sciences, Munster Technological University, Bishopstown, Cork, T12 P928, Ireland}
\author[0000-0002-7523-7366]{D.~Ribeiro}\affiliation{School of Physics and Astronomy, University of Minnesota, Minneapolis, MN 55455, USA}
\author{E.~Roache}\affiliation{Center for Astrophysics $|$ Harvard \& Smithsonian, Cambridge, MA 02138, USA}
\author[0000-0003-1387-8915]{I.~Sadeh}\affiliation{Deutsches Elektronen-Synchrotron DESY, Platanenallee 6, 15738 Zeuthen, Germany}
\author[0000-0002-3171-5039]{L.~Saha}\affiliation{Center for Astrophysics $|$ Harvard \& Smithsonian, Cambridge, MA 02138, USA}
\author[0009-0000-0295-8800]{H.~Salzmann}\affiliation{Santa Cruz Institute for Particle Physics and Department of Physics, University of California, Santa Cruz, CA 95064, USA}
\author{M.~Santander}\affiliation{Department of Physics and Astronomy, University of Alabama, Tuscaloosa, AL 35487, USA}
\author{G.~H.~Sembroski}\affiliation{Department of Physics and Astronomy, Purdue University, West Lafayette, IN 47907, USA}
\author[0009-0008-7331-7240]{S.~Tandon}\affiliation{Physics Department, Columbia University, New York, NY 10027, USA}
\author{J.~V.~Tucci}\affiliation{Department of Physics, Indiana University Indianapolis, Indianapolis, Indiana 46202, USA}
\author[0000-0003-2740-9714]{D.~A.~Williams}\affiliation{Santa Cruz Institute for Particle Physics and Department of Physics, University of California, Santa Cruz, CA 95064, USA}
\author[0000-0002-2730-2733]{S.~L.~Wong}\affiliation{Physics Department, McGill University, Montreal, QC H3A 2T8, Canada}
\author[0009-0001-6471-1405]{J.~Woo}\affiliation{Columbia Astrophysics Laboratory, Columbia University, New York, NY 10027, USA}
\author{T.~Yoshikoshi}\affiliation{Institute for Cosmic Ray Research, University of Tokyo, 5-1-5, Kashiwa-no-ha, Kashiwa, Chiba 277-8582, Japan}

\correspondingauthor{Tobias Kleiner}
\email{tobias.kleiner@desy.de}



\begin{abstract}
Microquasars such as SS~433 are considered potential contributors to cosmic rays up to the knee of the cosmic ray energy spectrum ($\sim4\,\mathrm{PeV}$), where a transition in the dominant acceleration processes is expected. The SS~433 system, located within the W50 supernova remnant, is a Galactic microquasar with relativistic jets interacting with the surrounding medium over parsec scales, providing an example for studying jet-driven particle acceleration.
A deep morphological and spectral study of SS~433 is performed using more than 150 hours of observations with VERITAS, sensitive to $\gamma$-ray energies $>100\,\mathrm{GeV}$. With an angular resolution better than \SI{0.1}{\degree}, extended TeV $\gamma$-ray emission is resolved from both the eastern and western jet lobes, located tens of parsecs from the central binary. The emission appears elongated along the jet axis and coincides with regions where the jets interact with the surrounding supernova remnant. No TeV emission is detected from the central binary, nor is significant emission observed between the central binary and the jet lobes. Phase-resolved analyses show no evidence for variability with orbital or precessional phase, supporting a steady emission scenario.
The observed morphology and spectra are consistent with scenarios where particles are accelerated in the lobes of the jets, possibly through shocks or alternative processes such as magnetic reconnection. The extended TeV emission from the jet lobes of SS~433 favors a leptonic origin in the VERITAS energy range, suggesting any hadronic acceleration is subdominant. The results offer valuable constraints on how microquasar jets may contribute to the Galactic cosmic-ray population toward the knee.
\end{abstract}




\section{Introduction}
\label{sec:intro}
Microquasars are compact binaries in which accreting compact objects launch relativistic jets, making them valuable laboratories for studying particle acceleration and feedback on the surrounding medium. SS~433, discovered in 1977 \citep{stephensonNewHalphaEmission1977}, is a well-studied Galactic microquasar and a unique example of a system with baryonic, precessing jets embedded in the supernova remnant (SNR) W50 \citep{marshallHighResolutionXRaySpectrum2002, margonEnormousPeriodicDoppler1979}. Evidence for the baryonic content of the jets is provided by a hot X-ray continuum with Doppler-shifted iron lines \citep{migliariIronEmissionLines2002} and optical emission from hydrogen and helium. Located at a distance of $\sim 5.5\,\mathrm{kpc}$, the binary system consists of a $>7\,M_\odot$ black hole accreting from a massive A-type donor star ($>19\,M_\odot$) \citep{bowlerSS433Two2018a, cherepashchukMassesComponentsSS4332018} and has an orbital period of 13 days. 
Jets with a velocity of $0.26\,c$ are launched perpendicular to the orbital plane and precess with a period of 162 days. Close to the binary, the jets are visible in radio for only a few precession cycles before fading. At distances of $\sim$\SI{25}{pc} from the central black hole, the jets re-emerge in X-rays and $\gamma$-rays, and ultimately terminate at distances from the black hole of $\sim$\SI{96}{pc} with a reduced velocity of $<0.04\,c$ due to interaction with the interstellar medium \citep{goodallProbingHistorySS2011}.
Multiwavelength observations have revealed a complex picture of the system. In X-rays, distinct emission regions in the jet lobes show spectral hardening along the jet axis \citep{safi-harbHardXRayEmission2022}, and polarization measurements indicate synchrotron emission in an ordered magnetic field aligned with the jets \citep{kaaretXRayPolarizationEastern2024}. At TeV energies, detections by HAWC, LHAASO and H.E.S.S. demonstrate efficient particle acceleration up to at least PeV energies \citep{abeysekaraVeryhighenergyParticleAcceleration2018,liGammarayHeartbeatPowered2020,caoFirstLHAASOCatalog2023,h.e.s.s.collaborationAccelerationTransportRelativistic2024,lhaasocollaborationUltrahighEnergyGammarayEmission2024}. HAWC first reported an excess of high-energy photons up to \SI{25}{TeV} \citep{abeysekaraVeryhighenergyParticleAcceleration2018}, followed by an updated analysis in \citet{alfaroSpectralStudyVeryhighenergy2024}. 
Fermi-LAT observations (\SI{100}{MeV}–\SI{300}{GeV}) revealed $\gamma$-ray emission from the source J1913+0515, located slightly north of the eastern jet lobe and off the jet axis. This emission may be associated with SS~433 and shows hints of precessional variability \citep{liGammarayHeartbeatPowered2020}.
LHAASO-KM2A detected very-high-energy TeV emission separated by $0.15^\circ$ from the eastern e1 region (see \citep{safi-harbROSATASCAObservations1997} for details on X-ray defined emission regions) at $\sim$50~TeV  \citep{caoFirstLHAASOCatalog2023}. H.E.S.S. later confirmed extended emission from $0.8\,\mathrm{TeV}$ to above $10\,\mathrm{TeV}$) from both lobes and revealed an energy-dependent morphology, with higher-energy emission concentrated closer to the central binary and lower-energy emission farther outward \citep{h.e.s.s.collaborationAccelerationTransportRelativistic2024}. A subsequent LHAASO study detected TeV emission from the jet lobes at 1–\SI{25}{TeV} and ultra-high-energy ($\geq$\SI{100}{TeV}) photons from a region closer to the central binary \citep{lhaasocollaborationUltrahighEnergyGammarayEmission2024}, challenging purely leptonic scenarios for the emission at the lobe as first discussed by HAWC. Despite these advances, questions remain concerning the detailed morphology of the TeV emission, the possible presence of central $\gamma$-ray emission, and flux variability linked to orbital or precessional phases.

During non-flaring periods, the kinetic luminosity of the jets is estimated to be 
\(L_{\mathrm{kin}}^{\mathrm{jet}} \sim 10^{39}\,\mathrm{erg\,s^{-1}}\) 
\citep{margonObservationsSS4331984}, which over the system’s lifetime corresponds to a total energy release of 
\(\sim 10^{51}\)–\(10^{52}\,\mathrm{erg}\). Although the majority of the accreted mass is expelled through powerful, super-Eddington disk winds 
\citep{perezm.SS433Circumbinary2010}, the kinetic energy of the jets defines the total energy available for particle acceleration within the jets. 
This provides an important constraint for distinguishing between leptonic and hadronic emission models, 
as purely hadronic scenarios can demand source power that approaches or even exceeds the total kinetic energy budget available for acceleration over the system’s lifetime, effectively ruling them out.

In this work, over 150 hours of VERITAS observations are analyzed to provide an independent and complementary view of SS~433 at TeV energies. With an angular resolution better than $0.1^\circ$, VERITAS enables sensitive morphological studies of the eastern and western jet lobes and searches for emission near the compact object. The temporal distribution of the observations allows phase-resolved searches for variability. Finally, the emission from the jet lobes is modeled under leptonic and hadronic scenarios, providing additional constraints on the acceleration and radiation processes.

\section{VERITAS DATA ANALYSIS}\label{sec:veritas_analysis}

The VERITAS (Very Energetic Radiation Imaging Telescope Array System) four-telescope imaging atmospheric Cherenkov telescope (IACT) array is located at the Fred Lawrence Whipple Observatory (FLWO) in southern Arizona (31 40N, 110 57W,  1.3km a.s.l.) \citep{weekesVERITASVeryEnergetic2002}. VERITAS observes $\gamma$-ray sources over the energy range from \SI{100}{GeV} up to $>$\SI{30}{TeV} with an energy resolution of 15-\SI{25}{\percent}, angular resolution of $<0.1\,^{\circ}$ (\SI{68}{\percent} containment radius) at \SI{1}{TeV} and a sensitivity to detect a source with \SI{1}{\percent} of the flux of the Crab Nebula within \SI{25}{\hour} \citep{parkPerformanceVERITASExperiment2016}.

\begin{figure}[ht]
	\includegraphics[width=1\linewidth]{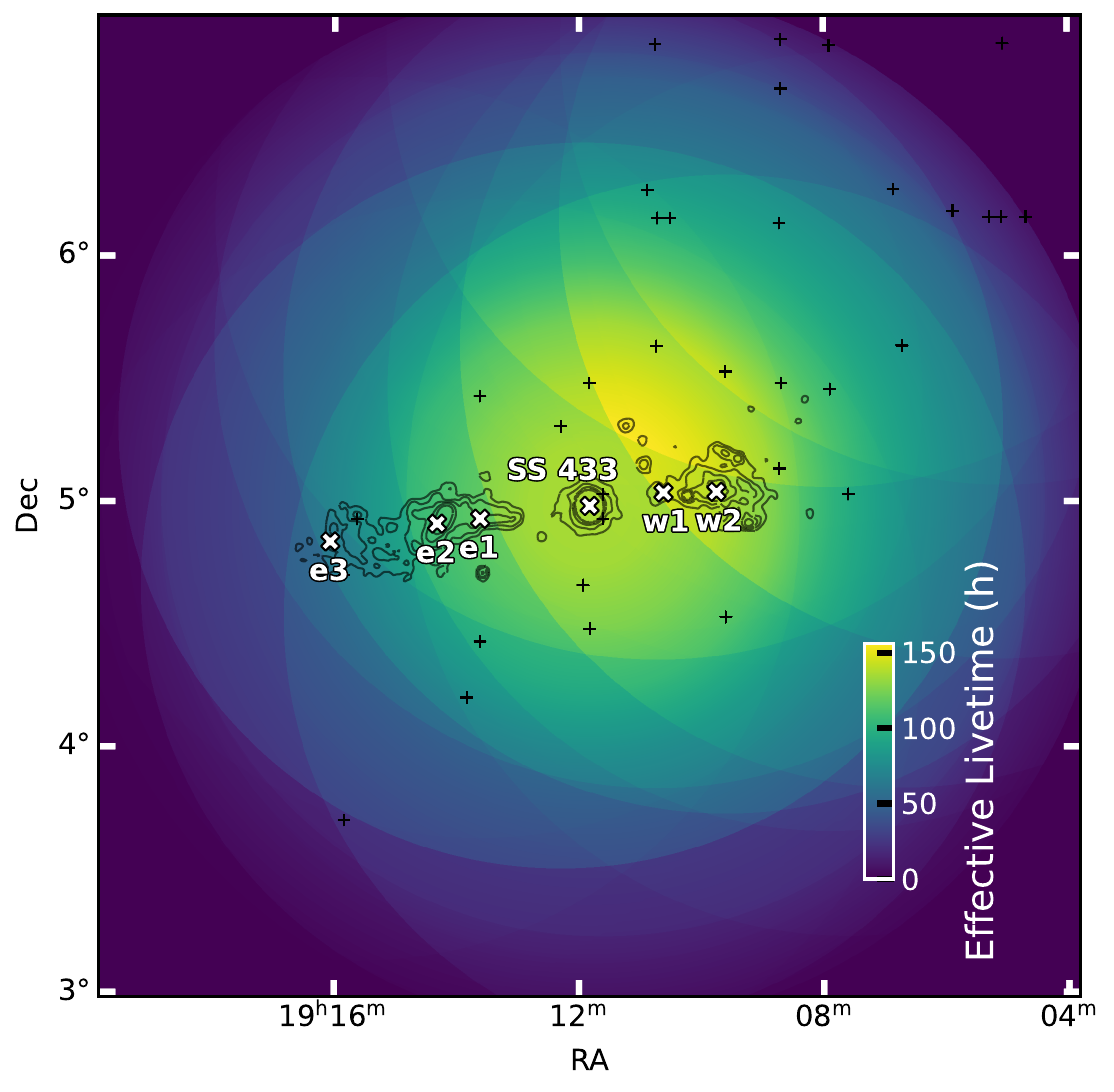}
	\centering
\caption[Acceptance corrected livetime]{SS 433 region acceptance corrected livetime map: Computed by dividing the exposure map by the VERITAS on axis effective area evaluated at an energy of \SI{1}{TeV}. The map is overlaid by black ROSAT X-ray contours \citep{brinkmannROSATObservations501996}. The VERITAS observation pointings are indicated by black markers (+) and the eastern (e1, e2, e3) and western (w1, w2) jet emission regions and the central binary (SS 433) are indicated by white crosses. The extended source MGRO J1908+06 is situated at the top right corner.}
  	\label{fig:ss433:ss433_acceptance_corrected_exposure}
\end{figure}

VERITAS has conducted observations of SS 433 for over a decade. This analysis includes data collected between 2009 and Sep 2023, with acceptance-corrected live-times of approximately \SI{100}{h} and \SI{150}{h} for the eastern and western jet lobes, respectively. Given the limited VERITAS field of view of \SI{3.5}{\degree} and the reduced sensitivity to extended sources arising from the limited camera area for background estimation, a dedicated pointing strategy is required. The live-time distribution and corresponding VERITAS pointings are shown in Figure~\ref{fig:ss433:ss433_acceptance_corrected_exposure}. A substantial fraction of the observations were designed to target the nearby $\gamma$-ray source MGRO~J1908+06, which, due to the proximity of the two objects, resulted in significant joint exposure on SS~433. The zenith-angle distribution of the observations is provided in Table~\ref{tab:zenith_angle}. The majority of runs are taken at low zenith angles of around \SI{30}{\degree} - \SI{40}{\degree} or less, minimizing potential systematic effects related to zenith dependence. In azimuth, the observations are primarily located in the range between $\sim100^{\circ}$–$250^{\circ}$ (from North in the clockwise direction).

\begin{table}[ht]
\centering
\begin{tabular}{ccccc}
\hline
\textbf{Zenith Range} & \textbf{$0^\circ$--$30^\circ$} & \textbf{$30^\circ$--$40^\circ$} & \textbf{$40^\circ$--$50^\circ$} & \textbf{$50^\circ$--$60^\circ$} \\ \hline
\textbf{Exposure [h]} & 107.8 & 58.3 & 17.3 & 6.5 \\ \hline
\end{tabular}
\caption{Zenith angle distribution of the observations with corresponding  exposure times.}
\label{tab:zenith_angle}
\end{table}

The low-level analysis is performed using the standard VERITAS analysis package Eventdisplay (v491) \citep{maierEventdisplayAnalysisReconstruction2017} up to DL2 data level, i.e., including calibration, image cleaning, and reconstruction of event parameters such as direction and energy \citep{nigroEvolutionDataFormats2021}. For $\gamma$-hadron separation, boosted decision tree (BDT) cuts optimized for hard-spectrum sources (spectral index $<2.5$) are applied \citep{krauseImprovedGammaHadron2017}. During the low level analysis, gain and total throughput corrections are applied \citep{adamsThroughputCalibrationVERITAS2022}. The resulting DL2 files are converted to DL3 format using the VERITAS (VEGAS \citep{coganVEGASVERITASGammaray2007a} and Eventdisplay) to DL3 Converter (v2dl3, v0.6.0) \citep{birdV2DL3VERITASVEGAS2022}. During conversion, the instrument response functions (IRFs), including effective area, energy dispersion, and point-spread function, are interpolated for the average observational conditions of each observation and include the dependence of the instrument response on offset relative to the pointing direction, known as full-enclosure analysis \citep{nigroEvolutionDataFormats2021}.

The high-level analysis is conducted with Gammapy (v1.3) \citep{aceroGammapyPythonToolbox2022} using a 3D maximum-likelihood approach. 
Background models are constructed from archival blank-field observations and describe the residual rate of $\gamma$-ray--like events (events surviving the $\gamma$--hadron separation cuts) as a function of field-of-view (FoV) camera coordinates and reconstructed energy. These models are generated in a sliding-window scheme in zenith (window size of \SI{10}{\degree}, steps of \SI{2}{\degree}) and separated into northern and southern azimuth regions, thereby accounting for variations in observing conditions and instrument epochs. Dependencies on offset, $\gamma$-hadron cuts, and instrument configuration are included.

The VERITAS systematic uncertainty budget affecting the reconstructed $\gamma$-ray fluxes and energies is well characterized in previous studies \citep{adamsThroughputCalibrationVERITAS2022}. Contributions include uncertainties in the photo-electron response and gain, pulse shape, low-gain channel modeling, pixel availability, mirror reflectivity, optical efficiency, point-spread function, atmospheric modeling, and the analysis and reconstruction pipelines. Overall, the combined systematic uncertainty on the flux is estimated to be $\sim$25\% and $\pm0.2$ on the spectral index for a source with a spectral index of $\approx 2.5$. These uncertainties, together with potential biases from the analysis method evaluated using independent mimic datasets (Appendix~\ref{app:systematic_bias}), are incorporated in the reported fluxes, ensuring that the total uncertainties reflect both statistical and systematic contributions.

To minimize systematic uncertainties from the analysis method, several selection criteria are applied. Observations are required to have a maximum pointing offset of \SI{1.8}{\degree} relative to the instrument pointing position. The energy threshold for each run is defined as the energy at which the effective area first reaches \SI{10}{\percent} of its maximum. In addition, a global lower energy threshold of \SI{0.8}{TeV} is applied to the
spectral analysis. At lower energies, threshold effects cause the residual background energy spectrum to deviate from a simple power-law shape. The analysis is therefore restricted to energies above the peak of the background energy distribution, where the background can be described by a power law. In this regime, the background model is corrected in both normalization and spectral slope. The upper energy limit of the stacked dataset is set to \SI{25}{TeV}, corresponding to the validity range of the background models used in this analysis.

\subsection{VERITAS Background Estimation Methods}

For the field-of-view background estimation \citep{bergeBackgroundModellingVeryhighenergy2007}, the observations are stacked after applying the above cuts. A power-law-based correction is applied to the stacked background, allowing the normalization and spectral tilt to vary freely. In Gammapy, the template background is then re-normalized across the entire FoV (outside of the exclusion mask) by fitting it to the dataset counts. The normalization is only performed if the minimum requirements on counts ($>0$) or predicted background events ($>0$) are satisfied, ensuring statistical robustness.

\begin{figure}
    \centering
    \includegraphics[width=1\linewidth]{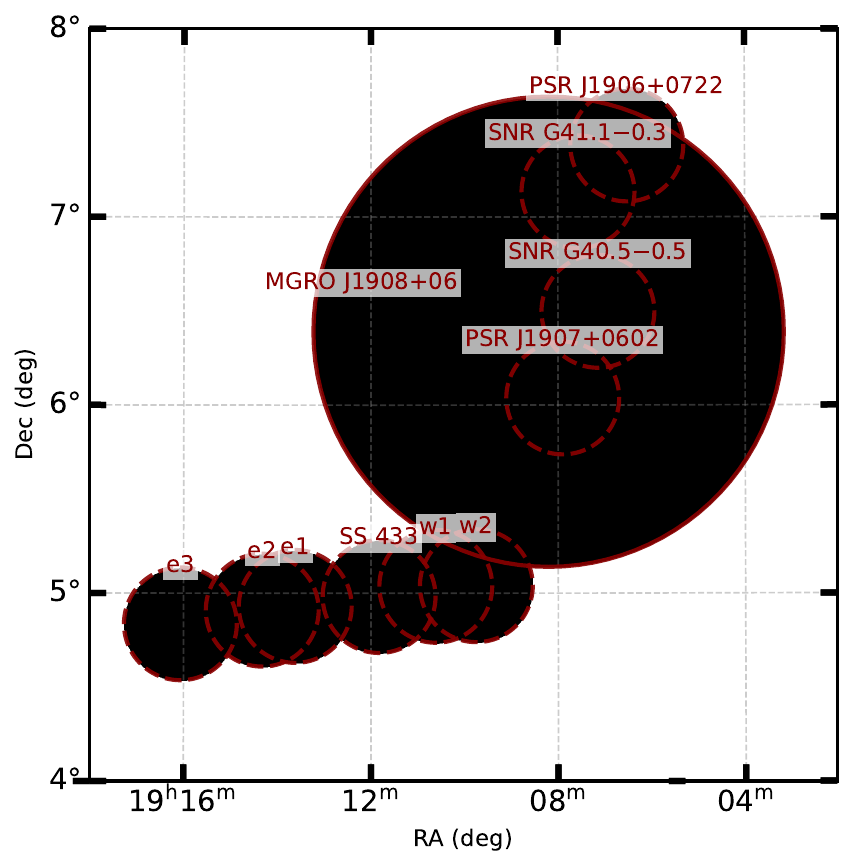}
    \caption{
        Exclusion mask used in the analysis. Red circles indicate regions around known sources and areas excluded from the analysis to avoid contamination.
        The large circle corresponds to MGRO~J1908+06, while smaller circles mark pulsars, SNRs, and the SS~433 jet regions (e1, e2, e3, w1, w2).
    }
    \label{fig:exclusion_mask}
\end{figure}

Circular exclusion regions are defined at the positions of known or candidate $\gamma$-ray sources to prevent contamination of the background model. In particular, a large exclusion region is applied around MGRO~J1908+06 (radius \SI{1.25}{\degree}), while smaller exclusion zones are placed at the positions of known pulsars and supernova remnants in the FoV (PSR~J1906+0722, PSR~J1907+0602, SNR G40.5-0.5, SNR G41.1-0.3; \SI{0.3}{\degree} radius), the central binary system of SS~433 (radius \SI{0.3}{\degree}), and the X-ray-defined eastern (e1, e2, e3) and western (w1, w2) jet emission regions (radius \SI{0.3}{\degree}). These exclusion radii are chosen based on the known spatial extent of $\gamma$-ray/X-ray emission in each region (see Figure~\ref{fig:exclusion_mask}).

In addition, a potential systematic bias stemming from the high level analysis method is analysed using mimic datasets for the regions of the jet lobes. The bias estimation is applied to energies above the established minimum energy threshold, and the expected systematic deviation in the flux estimates is predicted (see Appendix \ref{app:systematic_bias}).

\subsection{Modeling of MGRO~J1908+06 and the SS~433 Region}
\label{sec:modelling}

\begin{table}[htbp]
\centering
\begin{tabular}{@{}ccc@{}}
\toprule
\textbf{Type} & \textbf{Parameter} & \textbf{Value (Error)} \\ \midrule
\multirow{2}{*}{spectral} & $\Gamma$ & $2.37\pm0.03$ \\
 & $\phi_0$ & $(1.23\pm 0.04)\times10^{-11} 1 / (\mathrm{TeV}\,\mathrm{s}\,\mathrm{cm}^2) $ \\
 \midrule
\multirow{3}{*}{spatial} & lon & $287.01^\mathrm{\circ}\pm 0.01^\mathrm{\circ}$ \\
 & lat & $6.33^\mathrm{\circ}\pm 0.02^\mathrm{\circ}$ \\
 & $\sigma$ & $0.47^\mathrm{\circ}\pm 0.02^\mathrm{\circ}$ \\ \bottomrule
\end{tabular}
\caption[Best-fit parameters]{Best fit parameters of a symmetric Gaussian spatial combined with a power-law spectral model, derived for the MGRO~J1908+06 region in the energy range between \SI{0.8}{TeV} to \SI{25}{TeV} (with reference energy $E_{\mathrm{ref}}=1\,\mathrm{TeV}$).}
\label{tab:mgro_fit_parameters_3d}
\end{table}

Given the proximity of SS 433 to the bright extended source MGRO~J1908+06, careful modelling is crucial to avoid potential cross-contamination at the SS~433 jet lobes. The MGRO~J1908+06 exclusion regions are chosen to match those used in \citet{acharyyaMultiwavelengthInvestigationGRay2024a} and the morphological model is constructed following the same approach.
MGRO~J1908+06 is modeled using a symmetric 2D Gaussian spatial component combined with a power-law spectral model. 
The spectral model is given by:
\begin{equation}
\phi(E) = \phi_{\mathrm{0}} \left( \frac{E}{E_{\mathrm{ref}}} \right)^{-\Gamma}\,,
\end{equation}
where \(\Gamma\) is the spectral index, \(\phi_0\) is the amplitude, and \(E_{\text{ref}}\) is the reference energy. 

The spatial model is described with the parametrization:
\begin{equation}
\Phi(lon,lat) = \frac{1}{2\pi\sigma^2} \exp \left(-\frac{1}{2}\frac{\theta^2}{\sigma_{\mathrm{eff}}^2}\right)\,,
\end{equation}
with the width $\sigma$ and a spatial separation $\theta$ to the model center. In case of a 2D elliptical Gaussian for the modelling of the SS 433 jet lobes, the model is evaluated with an effective radius:
\begin{equation}
\sigma_{\mathrm{eff}}(\text{lon}, \text{lat}) = \sqrt{
    (\sigma_M \sin(\Delta \phi))^2 +
    (\sigma_m \cos(\Delta \phi))^2
}\,,
\end{equation}
where \(\sigma_M\) and \(\sigma_m\) denote the widths of the major and minor axes, respectively, and \(\Delta \phi\) is the angular difference between the position angle of the evaluation point and the position angle of the Gaussian model. Here, \(\phi\) defines the orientation of the Gaussian's major axis on the sky.

Constraints are imposed on the model parameters to ensure that all components remain physically consistent and spatially localized to their intended regions. To mitigate contamination from the nearby bright source MGRO~J1908+06, its spatial and spectral models are fit simultaneously with those of the SS 433 jet lobes within a joint likelihood framework. After obtaining the best-fit parameters for MGRO~J1908+06, its contribution is fixed and subtracted from the model cube, ensuring that the residual emission attributed to the SS 433 regions is not biased by spillover from the adjacent source.

In this analysis, the MGRO~J1908+06 region is modelled with a single symmetric Gaussian spatial and power-law spectral model. The resulting best-fit model encompasses the positions of the pulsar PSR~J1907+0602 \textbf{\citep{ackermannFERMILATSEARCHPULSAR2010, duvidovichRadioStudyExtended2020a}} and its likely associated PWN, as well as the northern $\gamma$-ray emission region associated with SNR~G40.5-0.5 and SNR~G41.1-0.3 \citep{acharyyaMultiwavelengthInvestigationGRay2024a}. No significant residual emission is observed in these regions after subtracting the MGRO~J1908+06 model from the data, indicating that the single extended model sufficiently accounts for the observed emission within the entire MGRO~J1908+06 region. Consequently, additional source components are not required to describe the emission from this area.

\subsection{VERITAS Data Analysis Results}

\begin{figure*}
	\includegraphics[width=0.49\linewidth]{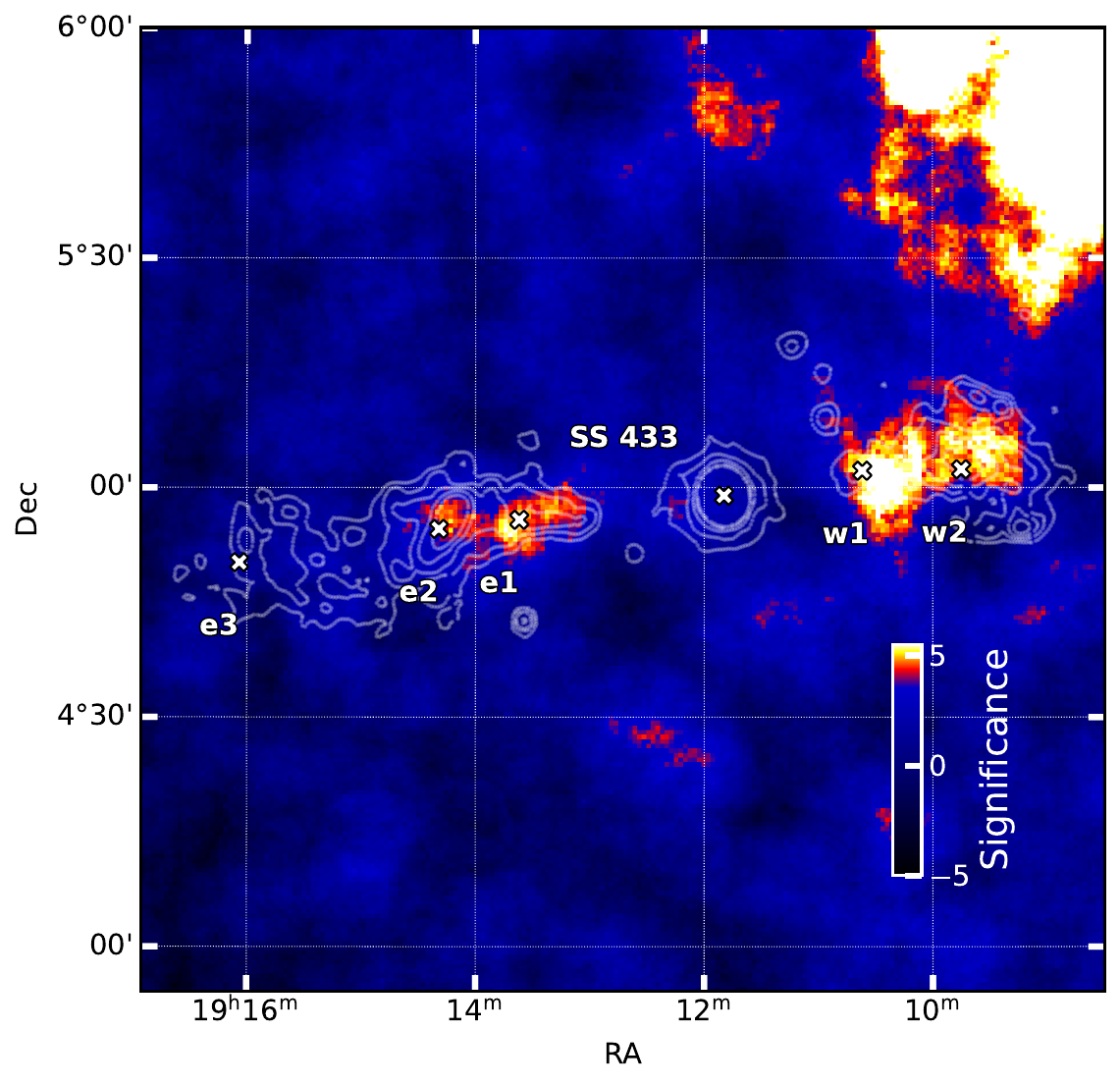}
	\includegraphics[width=0.49\linewidth]{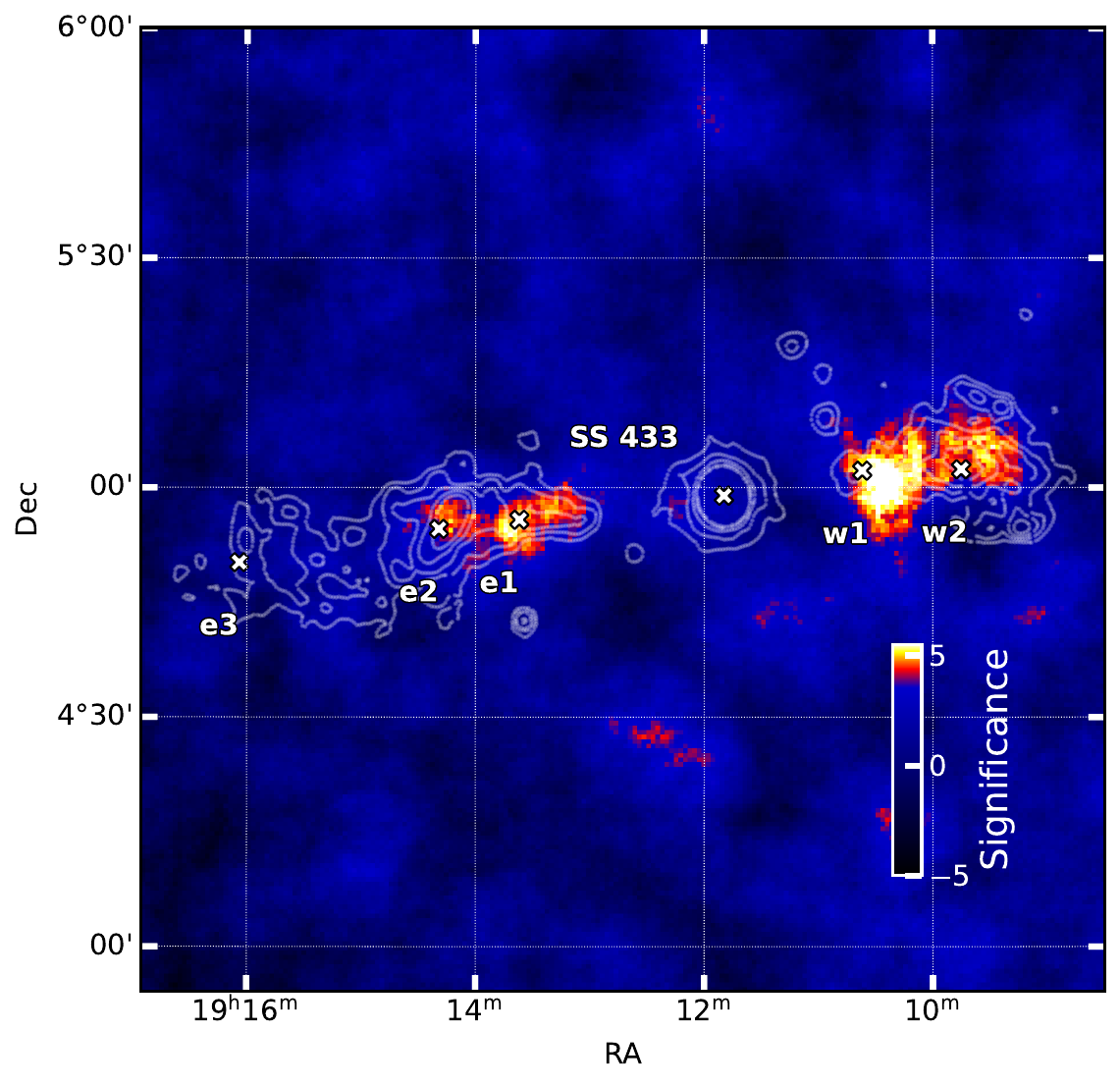}
	\centering
\caption[Significance Map]{SS 433 significance map before (left) and after (right) subtraction of the MGRO~J1908+06 best-fit model. White ROSAT X-ray contours \citep{brinkmannROSATObservations501996} are overlaid, and the eastern (e1, e2, e3) and western (w1, w2) jet emission regions, as well as the central binary (SS 433) position, are indicated by white crosses.}
  	\label{fig:ss433:ss433_sqrt_ts_map}
\end{figure*}

The significance map (Figure \ref{fig:ss433:ss433_sqrt_ts_map}) reveals distinct emission regions corresponding to SS 433's eastern and western jet lobes, hereafter referred to as VER~J1913+049 (east) and VER~J1910+050 (west).
Best-fit parameters for MGRO~J1908+06 obtained in this analysis (Table \ref{tab:mgro_fit_parameters_3d}) are consistent with the VERITAS measurements reported by \citet{acharyyaMultiwavelengthInvestigationGRay2024a}, with R.A., Decl., and spatial extent agreeing within statistical uncertainties. Minor differences between energy ranges are expected due to the different analysis setup and energy thresholds and do not affect the modelling of the SS~433 region.
The map is computed using the Cash counts statistic \citep{cashParameterEstimationAstronomy1979} in the energy range \SI{0.8}{TeV} - \SI{25}{TeV} and convolved with a top-hat kernel of \SI{0.1}{\degree} radius. 

Prior to subtracting the MGRO~J1908+06 model, the maximum significance reaches \SI{6.3}{\sigma} at the western jet lobe and \SI{5.5}{\sigma} at the eastern jet lobe. After subtraction of the MGRO~J1908+06 best-fit model, the significance decreases slightly to \SI{6.1}{\sigma} for the  western lobe, while remaining unchanged at \SI{5.5}{\sigma} for the eastern lobe.

\begin{figure}
	\includegraphics[width=1\linewidth]{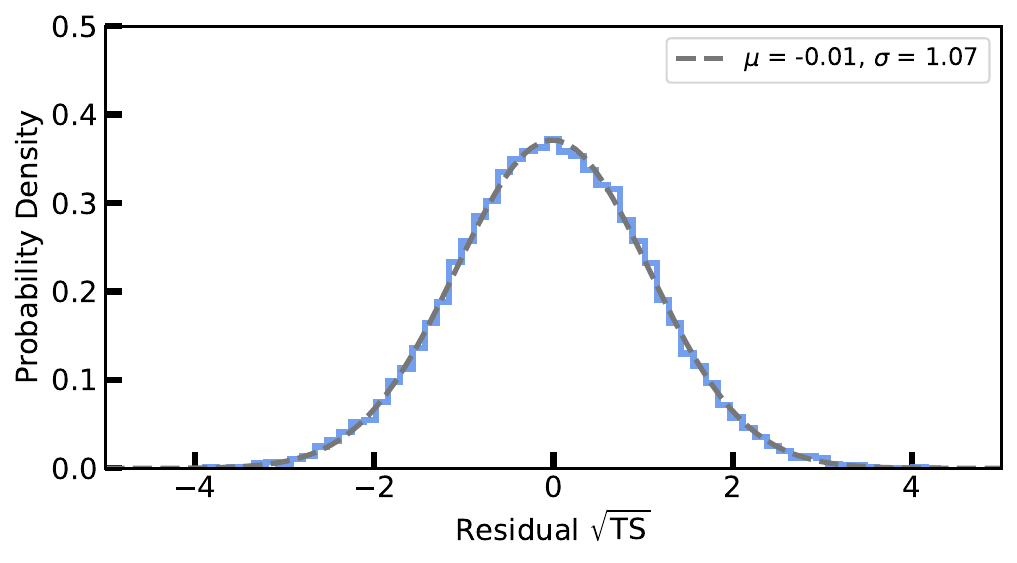}
	\centering
\caption[Residual TS]{Residual $\sqrt{\mathrm{TS}}$ distribution obtained after subtracting the best-fit models of SS 433 and MGRO J1908+06.}
  	\label{fig:ss433:residual_distribution_ss433_with_model_subtracted}
\end{figure}

To account for systematic effects in the significance estimation, a correction is applied based on the distribution of 
\(\sqrt{\mathrm{TS}}\) values across the FoV, where TS denotes the likelihood-based test statistic (see Figure \ref{fig:ss433:residual_distribution_ss433_with_model_subtracted}). 
The distribution is fitted with a Gaussian function, yielding a mean offset 
\(\mu\) and width \(\sigma\), which quantify deviations from an ideal standard normal distribution. 
The corrected significance values are obtained by re-shifting and re-scaling the significances according to
\begin{equation}
s_{\mathrm{corr}} = \frac{s_{\mathrm{pre}} - \mu}{\sigma}.
\end{equation}
For the measured parameters $\mu=-0.01$ and $\sigma=1.07$, the corrected significances are 
\(s_{\mathrm{corr}}^{\mathrm{W}} = 5.71\,\sigma\) and \(s_{\mathrm{corr}}^{\mathrm{E}} = 5.15\,\sigma\) for the western and eastern jet lobes, respectively.

The spatial properties of the $\gamma$-ray emission from the eastern and western jet lobes are examined using a model-based likelihood analysis, independent of the significance map results discussed above. This analysis provides a spectro-morphological characterization of the emission within a likelihood-ratio based framework. The emission from the eastern and western jet lobes is well described by elongated Gaussian spatial models combined with power-law spectral models. Extended morphologies are preferred over point-source models at $5.2\,\sigma$ (east) and $5.4\,\sigma$ (west), and over the null hypothesis at $6.0\,\sigma$ and $6.4\,\sigma$, respectively. A joint spectro-morphological fit yields an overall detection significance of $8.8\,\sigma$, with the elongated morphology favored over a symmetric Gaussian at $5.3\,\sigma$ jointly, and at $4.4\,\sigma$ (east) and $3.6\,\sigma$ (west) individually.

Fit constraints are applied to ensure robust convergence: positions are restricted to within \SI{0.3}{\degree} of the approximate emission centers of each lobe, defined by the observed maxima (e.g., between e1 and e2 for the eastern lobe), extensions are limited to a maximum of \SI{0.8}{\degree}, eccentricities ($e = \sqrt{1 - \sigma_m^2 / \sigma_M^2}$) are required to exceed 0.5, and the amplitude of the spectral model is constrained to $\phi_0 \le 10^{-12}\,\mathrm{cm^{-2}s^{-1}TeV^{-1}}$. Fits are repeated with varying initial parameters to confirm convergence to the global minimum.

\begin{figure}
	\includegraphics[width=1\linewidth]{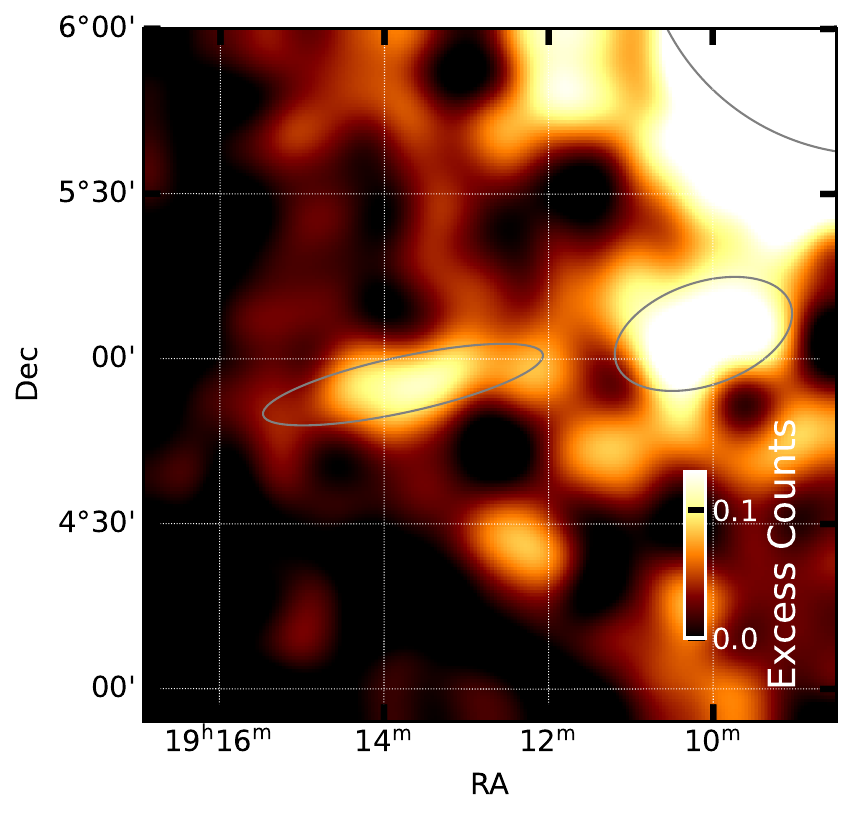}
	\centering
\caption[Residual Map]{SS 433 excess map with best fit model overlay ($1\,\mathrm{\sigma}$ contour).}
  	\label{fig:ss433:ss433_excess_map_blow_ups}
\end{figure}

The best-fit spatial parameters are summarized in Table~\ref{tab:ss433_fit_parameters_3d} and displayed in Figure \ref{fig:ss433:ss433_excess_map_blow_ups}. Both lobes are significantly elongated, with orientations aligned along the jet axis. 
As a cross-check, the residual maps after subtraction of the best-fit model have been inspected for residual emission, revealing no significant unmodeled features and confirming that the elongated Gaussian model adequately describes the observed $\gamma$-ray morphology.

The spectral properties of the eastern and western jet lobes are described by power-law models with indices $\Gamma = 2.62 \pm 0.20$ (east) and $2.53 \pm 0.16$ (west), consistent within statistical uncertainties (Table~\ref{tab:ss433_spectral_parameters}). The integrated fluxes above 0.8 TeV are comparable for both lobes, at the level of a few $10^{-13}\,\mathrm{cm}^{-2}\,\mathrm{s}^{-1}$.

\begin{figure}
	\includegraphics[width=1\linewidth]{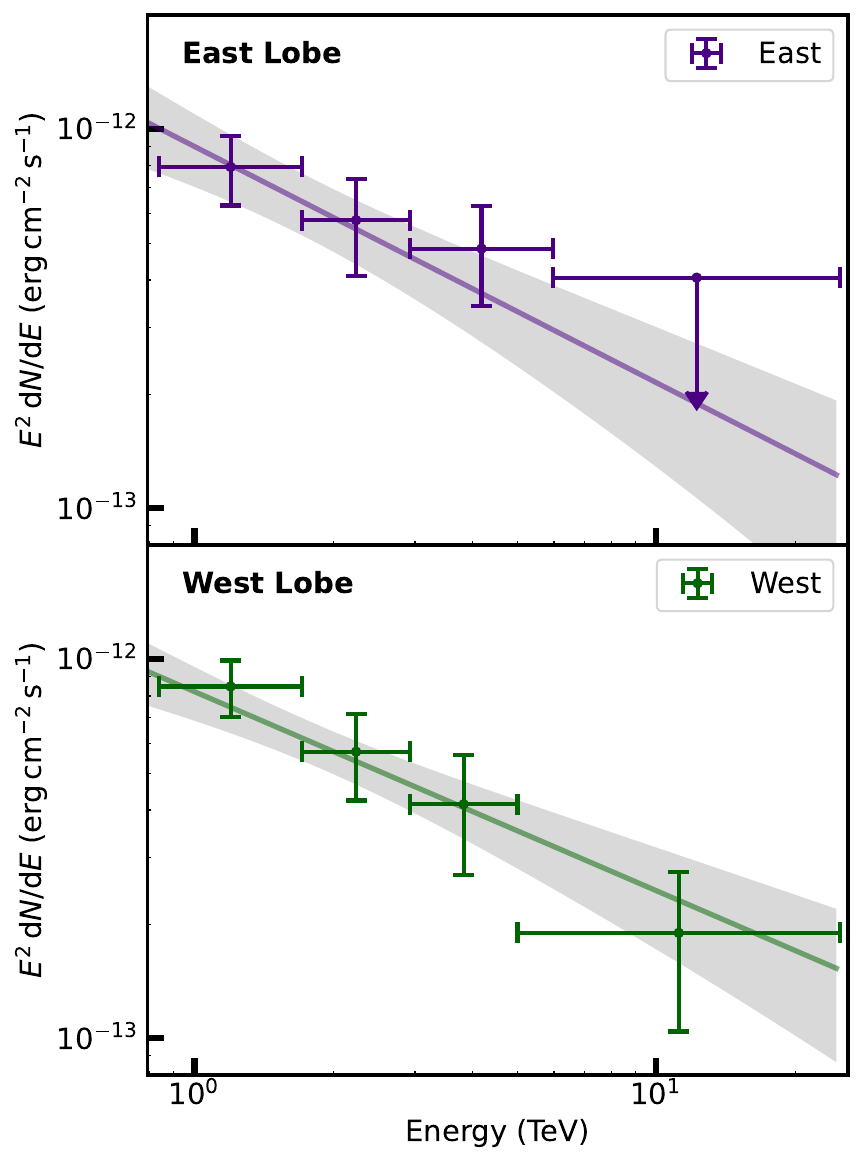}
	\centering
\caption[Flux points]{Flux points of SS~433 for the eastern and western jet lobes, computed from the best-fit model. Error bars represent $1\sigma$ statistical uncertainties, and the downward arrow denotes a $2\sigma$ upper limits.}
  	\label{fig:ss433:flux_comparison_east_west_jet_lobes}
\end{figure}

The flux points corresponding to the best fit models are depicted in Figure \ref{fig:ss433:flux_comparison_east_west_jet_lobes}. These are obtained by fitting the normalization of the spectral model independently in the four respective energy bins. The resulting normalization parameter quantifies the deviation of the flux from the reference model within that bin.

Motivated by previous H.E.S.S. results reporting an energy-dependent morphology, with higher-energy emission located closer to the central black hole and lower-energy emission extending further along the jet lobes, the energy-resolved significance maps were inspected for similar trends. No indication of an energy-dependent morphology is observed; however, the low significance in the individual energy bands limits the sensitivity to such effects.

\begin{table}[htbp]
\centering
\begin{tabular}{lcc}
\hline
\textbf{Parameter} & \textbf{East Lobe} & \textbf{West Lobe} \\
\hline
R.A. (deg) & $288.45 \pm 0.06$ & $287.53 \pm 0.04$ \\
Dec. (deg) & $4.92 \pm 0.02$ & $5.08 \pm 0.03$ \\
Major axis $\sigma$ (deg) & $0.28 \pm 0.06$ & $0.19 \pm 0.03$ \\
Minor axis $\sigma$ (deg) & $0.06 \pm 0.02$ & $0.11 \pm 0.02$ \\
Eccentricity & $0.98 \pm 0.01$ & $0.83 \pm 0.02$ \\
Angle (deg) & $103.3 \pm 6.4$ & $109.0 \pm 13.0$ \\
\hline
\end{tabular}
\caption[Elongated Gaussian best-fit parameters]{Best-fit parameters of elongated Gaussian models for the eastern (VER~J1913+049) and western (VER~J1910+050) jet lobes of SS~433.}
\label{tab:ss433_fit_parameters_3d}
\end{table}

\begin{table}[htbp]
\centering
\begin{tabular}{lccc}
\hline
\textbf{Lobe} & $\Gamma$ & $dN/dE$ & $F(E \geq 0.8\,\mathrm{TeV})$ \\
 & & [$10^{-14}$] & [$10^{-13}$] \\
 & & [TeV\(^{-1}\)\,cm\(^{-2}\)\,s\(^{-1}\)] & [cm\(^{-2}\)\,s\(^{-1}\)] \\
\hline
East & $2.62 \pm 0.20$ & $1.48 \pm 0.37$ & $4.53 \pm 1.60$ \\
West & $2.53 \pm 0.16$ & $1.55 \pm 0.28$ & $4.21 \pm 1.23$ \\
\hline
\end{tabular}
\caption[Spectral parameters]{Spectral parameters of the eastern and western SS~433 jet lobes from a power-law fit. $\Gamma$ is the spectral index, $dN/dE$ is the differential flux at 4\,TeV (in units of $10^{-14}$\,TeV\(^{-1}\)\,cm\(^{-2}\)\,s\(^{-1}\)), and $F(E \geq 0.8\,\mathrm{TeV})$ is the integrated flux above 0.8\,TeV (in units of $10^{-13}$\,cm\(^{-2}\)\,s\(^{-1}\)).}
\label{tab:ss433_spectral_parameters}
\end{table}

No significant $\gamma$-ray emission is detected at the black hole position or at the Fermi-LAT source Fermi J1913+0515. Upper limits (95\% C.L.) are summarized in Table~\ref{tab:ss433_upper_limits}, assuming power-law spectral indices of 2 (and 3). At the black hole position the derived upper limits correspond to $\approx0.25\,\%$ of the Crab flux at the same energy threshold. These limits complement the lobe flux measurements and are used in subsequent discussion of the central binary contribution.

\begin{table}[htbp]
\centering
\begin{tabular}{lc}
\hline
\textbf{Source} & \textbf{95\% C.L. Upper Limit} \\
\hline
SS 433 BH   & $7.5 (11) \times 10^{-14}\,\mathrm{cm^{-2}s^{-1}}$ \\
J1913+0515  & $4.8 (9.5) \times 10^{-14}\,\mathrm{cm^{-2}s^{-1}}$ \\
\hline
\end{tabular}
\caption[Point-like flux upper limits]{95\% confidence level upper limits on point-like TeV $\gamma$-ray emission at the position of the SS~433 black hole and Fermi-LAT source Fermi J1913+0515. Fluxes are computed assuming spectral indices of 2(3) and include systematic errors.}
\label{tab:ss433_upper_limits}
\end{table}

Potential flux variability of the SS~433 jet lobes is investigated along both precessional (162 days) and orbital (13 days) cycles. For phase calculations, the SS~433 ephemeris of reference, $T_{0} = \mathrm{JD}\,2443508.4098$, is adopted \citep{davydovSpectroscopicMonitoringSS2008}. During the orbital eclipse, occurring at $\mathrm{JD}\,2450023.62$, the optical star is thought to be located in front of the accretion disk \citep{cherepashchukINTEGRALObservationsSS4332013}.
Observations are grouped into precessional phases 0.0–0.5 (319 observations, $\sim 50 - 100\,\mathrm{h}$ at the jet lobes) and 0.5–1.0 (163 observations, $\sim 30 - 50\,\mathrm{h}$ at the jet lobes), and into five orbital phase bins corresponding to the ~13-day orbital period. The number of observations in the 5 orbital bins are 124, 89, 101, 81, and 87, with approximately similar exposure at the black hole region. For each bin, the best-fit model amplitudes are refitted in four energy intervals chosen to have approximately equal statistics, while all other model parameters are held fixed. Flux points are extracted both at the jet lobes and at the central binary position.

\begin{figure}
	\includegraphics[width=1\linewidth]{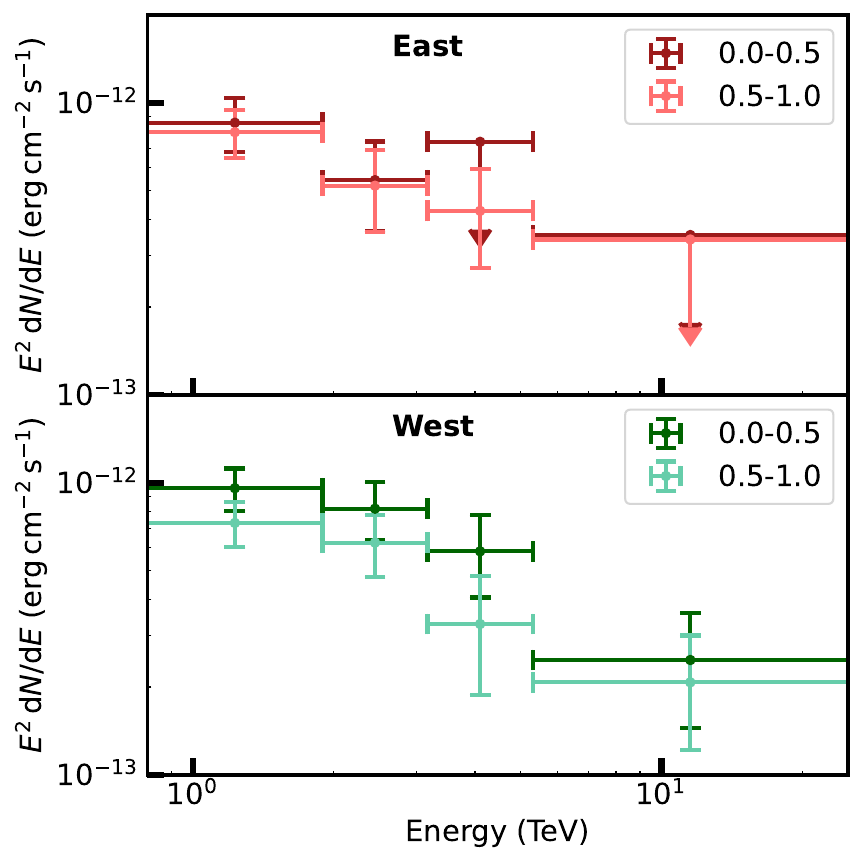}
	\centering
\caption[Flux in precessional phases]{Comparison of the $\gamma$-ray flux of SS~433 jet lobes across two precessional phase intervals (0.0–0.5 and 0.5–1.0). Top: eastern lobe; Bottom: western lobe.}
  	\label{fig:ss433:flux_comparison_precession_east_west_jet_lobes}
\end{figure}

\begin{figure}
	\includegraphics[width=1\linewidth]{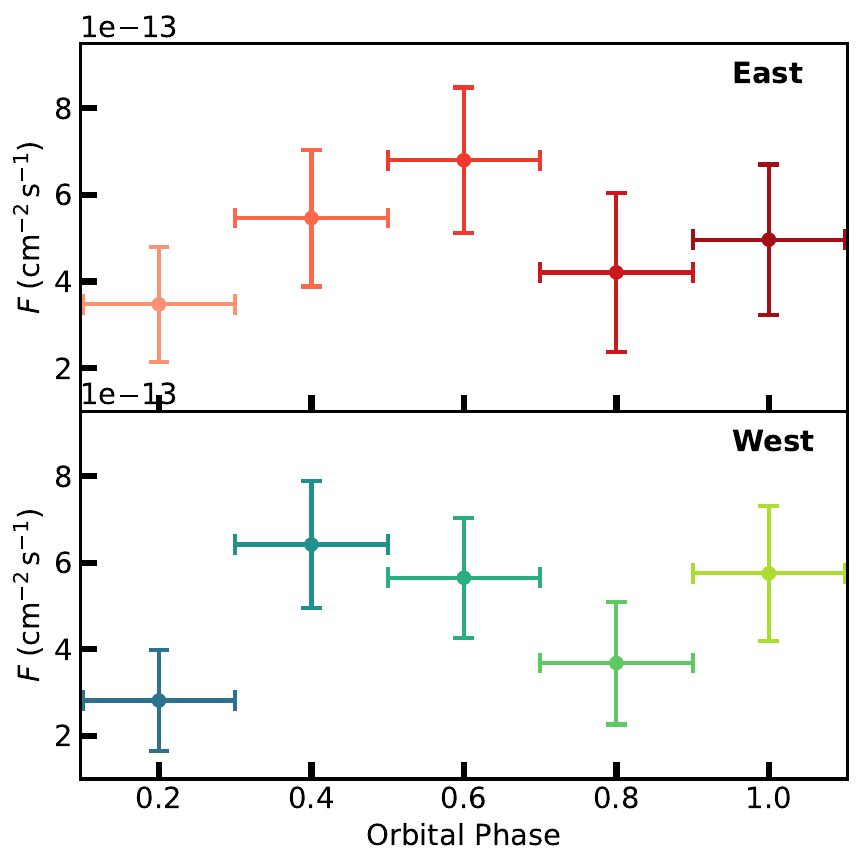}
	\centering
\caption[Flux in orbital phases]{Phase-resolved integrated $\gamma$-ray fluxes above \SI{0.8}{TeV} of the SS~433 jet lobes across five orbital phase bins (steps of 0.2). Top: eastern lobe; Bottom: western lobe.}
  	\label{fig:ss433:flux_vs_orbital_phase}
\end{figure}

No significant variability is observed in either precessional or orbital phase bins (Figures~\ref{fig:ss433:flux_comparison_precession_east_west_jet_lobes} and \ref{fig:ss433:flux_vs_orbital_phase}) at the jet lobes and the obtained flux values are consistent within error-bars. At the black hole position, 95\% C.L. upper limits on the integrated flux in all bins (orbital and precessional) are $\leq 8.65 \times 10^{-14}\,\mathrm{cm^{-2}\,s^{-1}}$, indicating that any phase-dependent modulation is well below sensitivity.

\section{Emission models for SS 433}
\label{sec:mw_sed_modeling}

In this section, a multi-wavelength spectral energy distribution (SED) for SS 433 is constructed, combining the VERITAS flux points, as well as previously reported flux values from radio, X-ray, high-energy (HE) and very-high-energy (VHE) observations. The eastern lobe is selected for this comparison due to the broader availability of flux measurements at other wavelengths and the lack of significant spectral differences between the eastern and western jet lobes observed with VERITAS. The comparison of these fluxes requires consideration of the spatial extent of the regions used for flux extraction at each wavelength, particularly for the X-ray and TeV bands, where the emission is spatially resolved.

In the radio regime, the flux upper limit provided by \SI{2695}{MHz} observations with the Effelsberg observatory \citep{geldzahlerContinuumObservationsSNR1980} is used. 
For the eastern region (spanning $~\sim 0.5^{\circ}\times 0.9^{\circ}$) encompassing the SS~433 jet termination area (regions e1–e2), a flux density of $5.9\pm1.6\,\mathrm{Jy}$ is reported, which is used here as an upper limit. 
In the soft X-ray range ($2-10\,\mathrm{keV}$), the XMM-Newton provides measurements of the eastern jet region in \citet{brinkmannXMMNewtonObservationsEastern2007} and more recently in \citet{safi-harbHardXRayEmission2022}.
The unabsorbed flux amounts to $(13.1\pm0.3)\times 10^{-12}\,\mathrm{erg}\,\mathrm{cm}^{-2}\mathrm{s}^{-1}$ in the energy range of $0.5\text{--}10\,\mathrm{keV}$.

The soft X-ray emission is taken from the \textit{Full} region defined in \citet{safi-harbHardXRayEmission2022} (see their Figure 2). This region comprises the three NuSTAR-defined subregions (“Head,” “Cone,” and “Diffuse”), which together represent the principal X-ray emitting structures of the inner eastern lobe, as well as the lenticular feature embedded within the larger e2 structure. The \textit{Full} region therefore includes the bulk of the hard and soft X-ray emission from the inner eastern lobe and broadly coincides with the radio e1–e2 emission region. It extends inward to a projected distance of approximately \SI{29}{pc} from the central binary and spans roughly $10'$ × $3'$ on the sky.


In the hard X-ray regime, NuSTAR provides measurements ($3-30\,\mathrm{keV}$) for the \textit{Full} region of the jet in \citet{safi-harbHardXRayEmission2022}. 
In the HE $\gamma$-ray range, a flux point at the e1 position was reported by Fermi-LAT in a joint analysis with HAWC \citep{fangGeVTeVCounterparts2020}.
The nearby Fermi source J1913+0515, whose association remains uncertain, was treated as background in that study. The upper limits derived in the same work are also included in the modelling.
In the VHE range, flux points from HAWC \citep{abeysekaraVeryhighenergyParticleAcceleration2018} correspond to emission centered on the e1–e2 region, with a point-spread function of $\gtrsim0.2^{\circ}$ above \SI{10}{TeV} \citep{abeysekara2HWCHAWCObservatory2017a}. The H.E.S.S. flux points correspond to an elliptical region encompassing both the e1 and e2 regions, with semi-major and semi-minor axes of $0.2055^{\circ}\pm0.035^{\circ}$ and $0.0445^{\circ}\pm0.014 ^{\circ}$, respectively \citep{h.e.s.s.collaborationAccelerationTransportRelativistic2024}, in good coincidence with the VERITAS best-fit region. The LHAASO flux points \citep{lhaasocollaborationUltrahighEnergyGammarayEmission2024} were obtained by fitting two point-like sources associated with the eastern and western lobes of SS~433. The best-fit positions were found toward the base of the jet lobes, and, given the relatively large point-spread function of LHAASO ($\gtrsim 0.5^{\circ}$ between $1-25\,\mathrm{TeV}$ and above), the fitted regions encompass a large fraction of the lobe emission.

\subsection{Leptonic SED Model}
\label{sec:mw_sed_modeling_leptonic}

\begin{figure*}
	\includegraphics[width=1\linewidth]{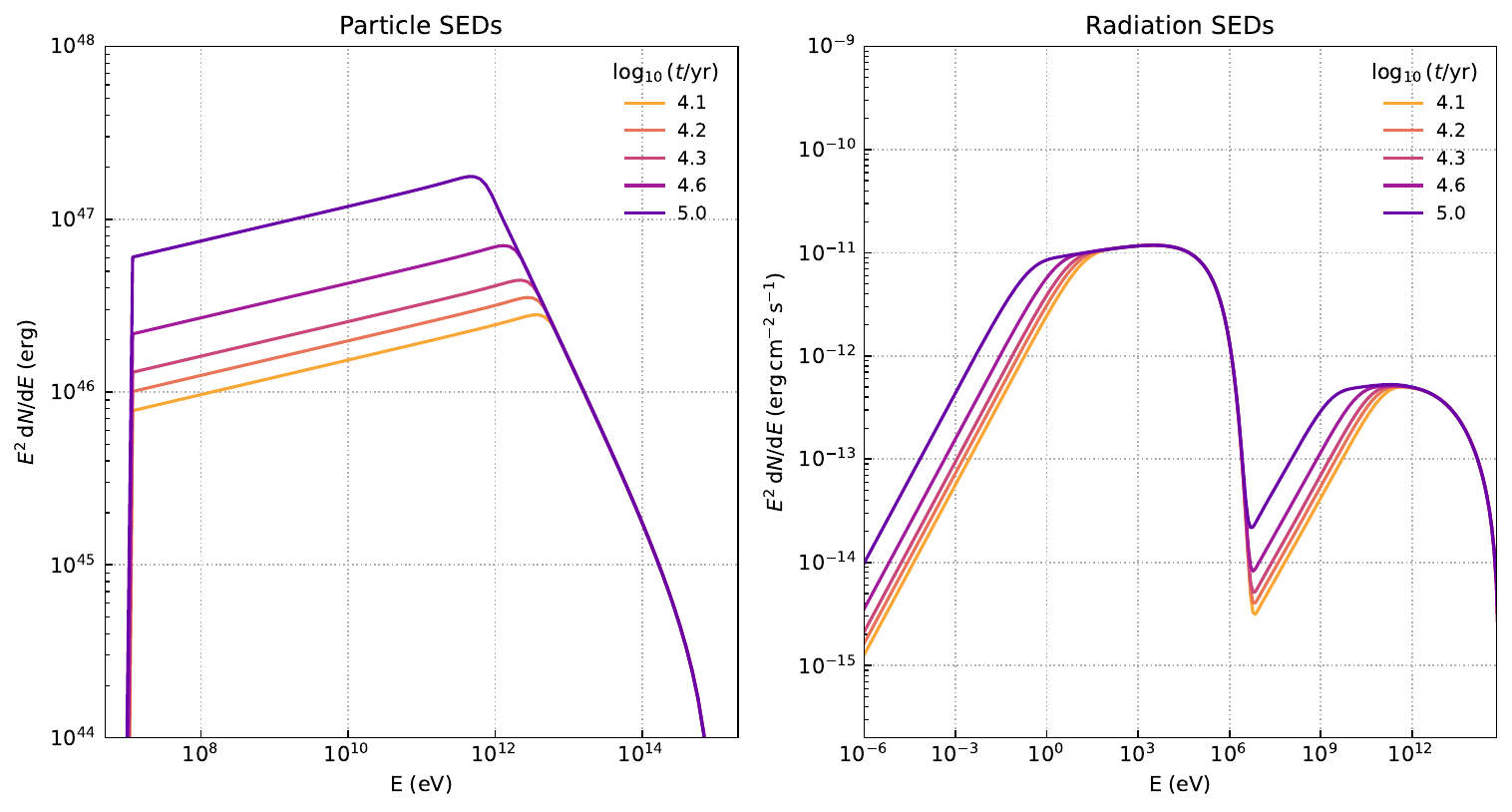}
	\centering
\caption[Time-evolved electron and radiation spectra]{Calculations of the time evolution of the electron spectral energy densities and the corresponding radiation SEDs, shown for five logarithmically spaced time steps between $10^4$ and $10^5$~years. The radiation components include synchrotron and inverse Compton (IC) emission on ambient photon fields, with a negligible SSC contribution. Electrons are injected with a power-law spectrum of index $\alpha = 1.9$ between \SI{10}{MeV} and \SI{1}{PeV}. The illustrated model assumes fixed environmental parameters representative of the computed scenario: CMB photon field with $u_{\mathrm{CMB}} = 0.261\,\mathrm{eV\,cm^{-3}}$ and $T = 2.7\,\mathrm{K}$, an FIR photon field with $u_{\mathrm{FIR}} = 0.3\,\mathrm{eV\,cm^{-3}}$ and $T = 20\,\mathrm{K}$, magnetic field $B = 14\,\mu\mathrm{G}$, and an injection power of $10^{36}\,\mathrm{erg\,s^{-1}}$.}
  	\label{fig:ss433:particles_static_timeseries}
\end{figure*}

\begin{figure*}
	\includegraphics[width=0.7\linewidth]{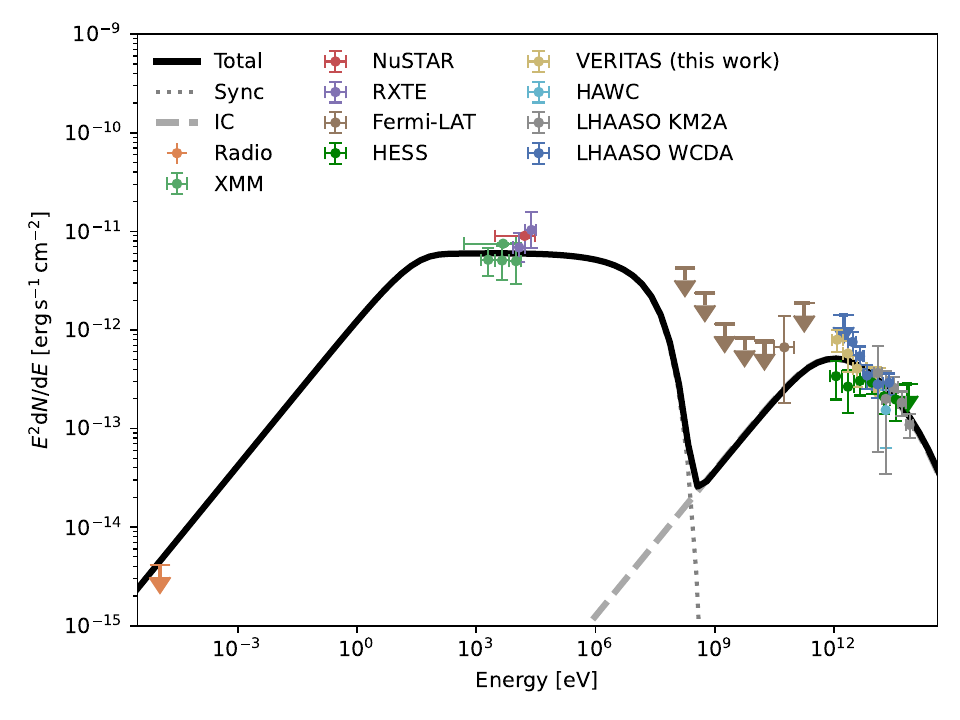}
	\centering
    \caption[Leptonic SED model comprising of Synchrotron and inverse Compton on the CMB, FIR using a exponential-broken power-law injection spectrum]{Multiwavelength spectral energy distribution of SS 433 eastern emission region, shown with a leptonic model. The observations are from radio \citep{geldzahlerContinuumObservationsSNR1980}, soft X-ray \citep{brinkmannXMMNewtonObservationsEastern2007}, hard X-ray \citep{safi-harbHardXRayEmission2022,safi-harbRossiXRayTiming1999}, HE \citep{fangGeVTeVCounterparts2020} and VHE VERITAS, H.E.S.S. \citep{h.e.s.s.collaborationAccelerationTransportRelativistic2024}, HAWC \citep{abeysekaraVeryhighenergyParticleAcceleration2018} and LHAASO \citep{lhaasocollaborationUltrahighEnergyGammarayEmission2024}. The VERITAS flux points include statistical and systematic errors. Electrons are injected with a power-law spectrum and they produce synchrotron radiation in the ambient magnetic field of the eastern jet emission region and high energetic photons via IC upscattering on the CMB and FIR photons. The spectrum corresponds to a source age of $5\cdot10^{4}\,\mathrm{yr}$.}
  	\label{fig:ss433:SS433_SynIC_gamera_evolved_pl_SED}
\end{figure*}

The flux points derived from this analysis are used to construct a multi-wavelength SED model.
These are combined with flux upper limits from radio and data points from soft and hard X-rays, as well as HE, VHE and UHE $\gamma$-rays.

\begin{table*}
\centering
\caption{
Parameters adopted in the leptonic SED model used for the \texttt{GAMERA} simulations.
The table indicates whether each parameter was fixed or computed during the modeling.
}
\begin{tabular}{llll}
\hline
\textbf{Parameter} & \textbf{Value} & \textbf{Description} & \textbf{Type} \\
\hline
$T_\mathrm{FIR}$ & 20 K & Temperature of far-infrared photon field & Fixed \\
$u_\mathrm{FIR}$ & $0.3\,\mathrm{eV\,cm^{-3}}$ & Energy density of FIR photon field & Fixed \\
$T_\mathrm{CMB}$ & 2.7 K & Temperature of CMB photon field & Fixed \\
$u_\mathrm{CMB}$ & $0.26\,\mathrm{eV\,cm^{-3}}$ & Energy density of CMB photon field & Fixed \\
Distance & 5.5 kpc & Distance to SS~433 & Fixed \\
$R$ & 3 pc & Radius of the electron injection region & Fixed \\
$E_\mathrm{tot}$ & — & Total energy of injected electrons & Computed \\
$\alpha$ & — & Spectral index of the injection spectrum & Computed \\
$B$ & — & Magnetic field strength & Computed \\
Age & — & Source age & Computed \\
\hline
\end{tabular}
\label{tab:sed_parameters}
\end{table*}

The leptonic SED model is parameterized by the total energy of injected electrons ($E_\mathrm{tot}$), the spectral index of the power-law injection spectrum ($\alpha$), and the magnetic field strength ($B$) in the emission region. Inverse Compton scattering is computed on both the cosmic microwave background (CMB) and a far-infrared photon field.
The presence of a significant FIR photon field around the eastern lobe of SS~433 is uncertain. No indications of dense material in this region have been reported \citep{fabrikaJetsSupercriticalAccretion2004}, whereas mid-infrared emission, likely of synchrotron origin, has been observed at the western lobe, coinciding with radio structures \citep{fuchsMidinfraredObservationsGRS2001}. A possible FIR component in the eastern region has been suggested to account for the enhanced flux observed at GeV energies relative to HAWC measurements \citep{fangGeVTeVCounterparts2020}. To quantify its potential effect, a FIR photon field with an energy density of $0.3\,\mathrm{eV\,cm^{-3}}$ and a black-body temperature of \SI{20}{K} has been included, following the prescription of \citet{vernettoAbsorptionVeryHigh2016} as adopted by \citet{fangGeVTeVCounterparts2020}. The age of the system and the size of the particle injection region are included as model parameters, as they determine the synchrotron photon field available for synchrotron self-Compton emission. Table~\ref{tab:sed_parameters} summarizes the parameters used in the simulations, highlighting which are fixed and which are varied in the procedure.

The spectral evolution of the electron population and the associated radiation SEDs are modelled using the \texttt{GAMERA} package \citep{hahnGAMERASourceModeling2022}. During computation, $E_\mathrm{tot}$, $\alpha$, and $B$ are adjusted to reproduce the observed SED, while the photon field parameters, distance, and region size are held fixed. 
The electron spectrum is evolved in time assuming constant injection and a finite source age, and the resulting electron distribution is used to compute synchrotron and inverse Compton radiation. The time-evolved electron spectra and corresponding radiation SEDs are depicted in Figure~\ref{fig:ss433:particles_static_timeseries}.

Electrons are injected into a spherical region of $6\,\mathrm{pc}$ diameter, corresponding approximately to the size of the X-ray emission region. The precise choice of region size does not significantly affect the modelled SED, since SSC contributes only minimally in this source. The spectral evolution accounts for radiation losses due to synchrotron emission, bremsstrahlung, ionization, and inverse Compton processes. The injection spectrum is assumed to be a power law, with parameters remaining constant during the time evolution. The SED is modelled for source ages ranging from \SI{10}{\kilo\year} to \SI{100}{\kilo\year}.
The resulting SED for a source age of \SI{50}{\kilo\year} is shown in Figure~\ref{fig:ss433:SS433_SynIC_gamera_evolved_pl_SED}. In this model, the total energy in electrons amounts to $\sim 9\times10^{47}\,\mathrm{erg}$, the magnetic field strength is $\sim13\,\mu\mathrm{G}$, and the electron injection spectrum power-law index is $\alpha \simeq 1.9$.

The magnetic field strength derived from the \texttt{GAMERA} simulations remains stable across the explored source ages. 
For assumed ages of 10, 50, and 100~kyr, the medians of the posterior distributions correspond to magnetic field strengths of approximately $14$--$17~\mu\mathrm{G}$, with spectral indices ranging between 1.9 and 2.2.

At a magnetic field of $B = 15\,\mu\mathrm{G}$, electrons with energies between $10^{12}$ and $10^{15}\,\mathrm{eV}$ are primarily cooled by synchrotron radiation, with inverse-Compton (IC) losses remaining subdominant. IC cooling would only dominate if the magnetic field were below $5\,\mu\mathrm{G}$. The relative importance of synchrotron versus IC losses determines the shape of the high-energy electron spectrum and the resulting photon emission. Above a few~TeV, the Klein--Nishina effect suppresses IC scattering, rendering synchrotron the dominant cooling process. At the highest energies ($\sim 10^{15}\,\mathrm{eV}$), electrons lose energy via synchrotron emission on timescales of $\sim 100$~years, whereas IC losses occur over several $10^{4}$~years.

Model degeneracies were examined by sampling the posterior distributions of the total injected electron energy ($E_\mathrm{tot}$), the magnetic field strength ($B$), and the spectral index of the injection spectrum ($\alpha$). 

From the posterior samples, uncertainties in $E_\mathrm{tot}$ were propagated over source ages between 10 and 100~kyr, yielding $E_\mathrm{tot} \simeq (0.3$--$1.4)\times10^{48}\,\mathrm{erg}$. When compared to the canonical jet kinetic power of SS~433 ($\sim 10^{39}\,\mathrm{erg\,s^{-1}}$), the required lepton injection corresponds to $\lesssim 0.1$ percent of the available jet power for the investigated source ages, supporting the plausibility of a leptonic emission scenario.

\subsection{Proton Density Estimates in the SS 433 Lobes}
\label{sec:proton_density}
Accurate proton-density estimates are essential for evaluating hadronic scenarios. To constrain the density of the ambient gas in the W50 cavity, into which the SS\,433 jets propagate, CO and HI data from regions spatially coincident with the observed $\gamma$-ray emission are used. Column densities are derived in selected velocity bands based on the HI4PI full-sky HI survey \citep{bekhtiHI4PIFullskySurvey2016} and MWISP CO observations \citep{suMilkyWayImaging2019}, providing estimates of the target material available for interactions with relativistic protons in the extended lobes.

The MWISP $^{12}\mathrm{CO}$ spectroscopic data is converted using the $X_{^{12}\mathrm{CO}}$ factor from \citet{bolattoCOtoH2ConversionFactor2013} to compute the molecular hydrogen column density $N_{\mathrm{H2}}$.
Likewise the HI4PI full sky HI survey data \citep{bekhtiHI4PIFullskySurvey2016} is converted to the atomic hydrogen column density
\citep{bekhtiHI4PIFullskySurvey2016}:
\begin{equation}
N_{\mathrm{HI}} (\mathrm{cm}^{-2}) = 1.823\cdot 10^{18}\int dv T_{\mathrm{B}}(v)\,,
\end{equation}
where $T_{\mathrm{B}}(v)$ is the brightness temperature profile in units of $\left(\mathrm{K}\,\mathrm{km}\,\mathrm{s}^{-1}\right)$.

\begin{figure*}
\includegraphics[width=1\linewidth]{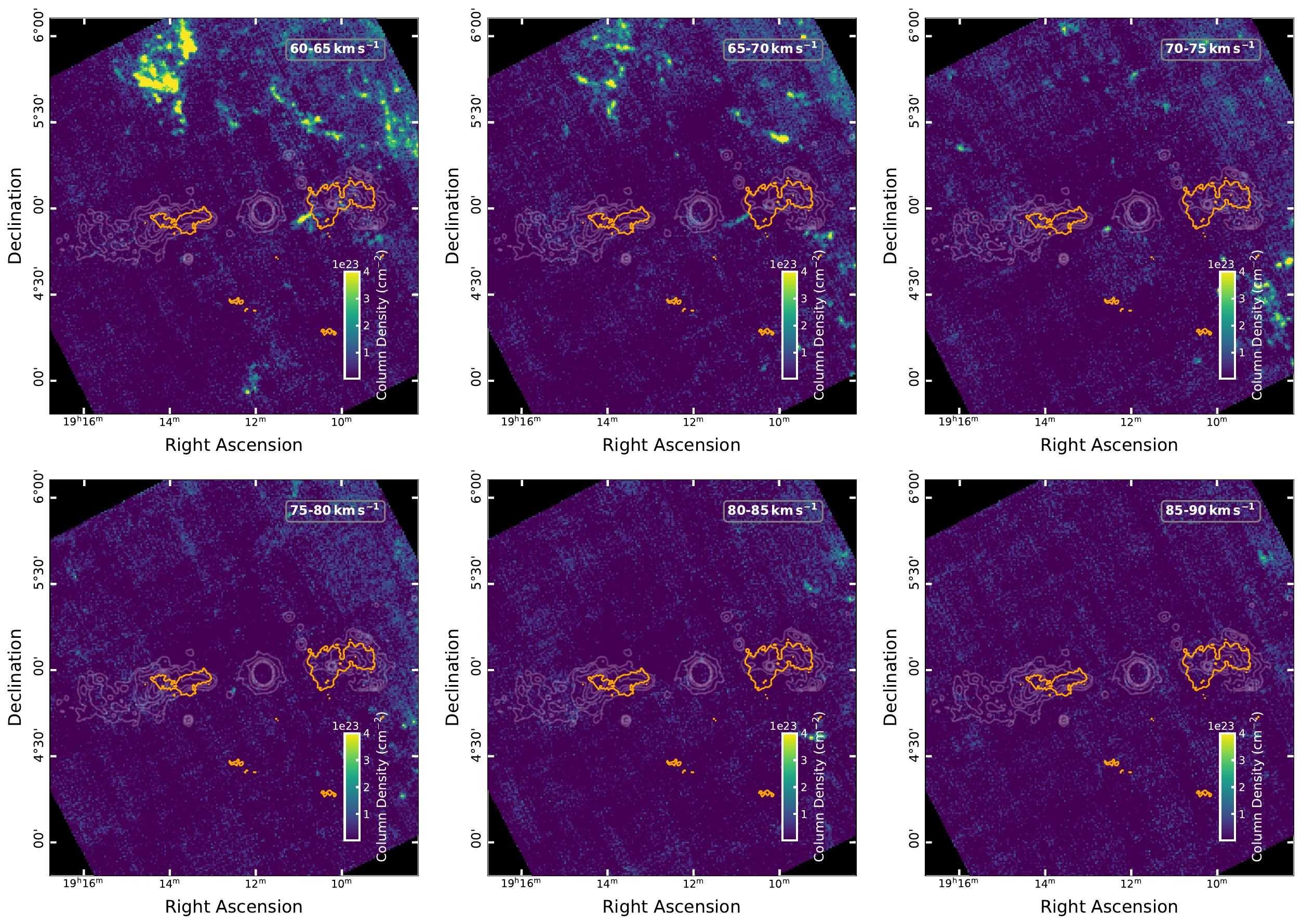}
\caption[$^{12}$CO column densities]{Column densities ($\mathrm{cm}^{-1}$) for spectroscopic data in the velocity bands $(60-65)\,\mathrm{km}\,\mathrm{s}^{-1}$, $(65-70)\,\mathrm{km}\,\mathrm{s}^{-1}$, $(70-75)\,\mathrm{km}\,\mathrm{s}^{-1}$, $(75-80)\,\mathrm{km}\,\mathrm{s}^{-1}$, $(80-85)\,\mathrm{km}\,\mathrm{s}^{-1}$ and $(85-90)\,\mathrm{km}\,\mathrm{s}^{-1}$ from $^{12}$CO. VERITAS \SI{4}{\sigma} contours in orange are overlaid on ROSAT X-ray contours in grey \citep{brinkmannROSATObservations501996}. Data by the MWISP project \citep{suMilkyWayImaging2019}.
}
\label{fig:ss433:ss433_skymap_CO_overview_mwisp}
\end{figure*}
Integrated column densities for several distance-velocity slices are shown in Figure \ref{fig:ss433:ss433_skymap_CO_overview_mwisp}. For further analysis, the velocity range $(70-80)\,\mathrm{km\,s^{-1}}$ is considered, corresponding to a distance of $\sim 4.6$–\SI{5.7}{kpc} according to the Milky Way rotation curve model of \citet{bhattacharjeeROTATIONCURVEMILKY2014}. Additionally, data from MWISP and the Arecibo Observatory \citet{suLargescaleInterstellarMedium2018} suggest an association with gas at $77\pm5\,\mathrm{km}\mathrm{s}^{-1}$, corresponding to a distance of $4.9\pm0.4\,\mathrm{kpc}$.
The derived column density in the MWISP dataset is up to $2.5\cdot 10^{22}\,\mathrm{cm}^{-2}$ in the eastern jet lobe and $9\cdot 10^{22}\,\mathrm{cm}^{-2}$ in the western lobe with maxima located in proximity to the e1 and w1 regions, respectively. The corresponding proton target densities in the eastern jet lobe region range between $n_{\mathrm{p}}=0.15$ and $n_{\mathrm{p}}=1.2\,\mathrm{cm}^{-3}$.

\subsection{Hadronic SED Model}
\label{sec:mw_sed_modeling_hadronic}

Based on the estimates in section \ref{sec:proton_density}, hadronic $\gamma$-ray emission is modeled at the eastern  jet lobe using the computational package Naima \citep{zabalzaNaimaPythonPackage2015}, assuming a minimum (maximum) proton target density as derived in that section of $n_{\mathrm{p}}=0.15\,(1.2)\,\mathrm{cm}^{-3}$.

A hadronic emission model is applied to account for the presence of protons in the jets of SS 433. Observations of line spectra indicate that the jets contain both protons and heavier nuclei \citep{migliariIronEmissionLines2002}. Charged protons emit synchrotron radiation in the presence of magnetic fields and interact with photon fields, leading to photo-meson production. However, the dominant $\gamma$-ray production mechanism in a hadronic scenario is proton-proton collisions, which result in the creation of neutral pions ($\pi^0$) that decay into $\gamma$-rays. The parametrization from \citet{kafexhiuParametrizationGammarayProduction2014} is used with the Pythia8 \citep{sjostrandIntroductionPYTHIA822015} high energy model to compute the resulting radiation spectrum from these hadronic processes.

The parent proton distribution follows an exponential cut-off power-law. Synchrotron emission is modeled with an exponential broken power-law spectrum. The minimum proton energy is set to \SI{1.22}{GeV}. Markov Chain Monte Carlo (MCMC) sampling is constrained by uniform priors: $A\geq 0$ for amplitudes, $1\leq\alpha_i\leq4$ for spectral indices, and $0\leq B\leq 30 \,\mathrm{\mu G}$ for the magnetic field.

\begin{figure}
    \centering
    \includegraphics[width=1\linewidth]{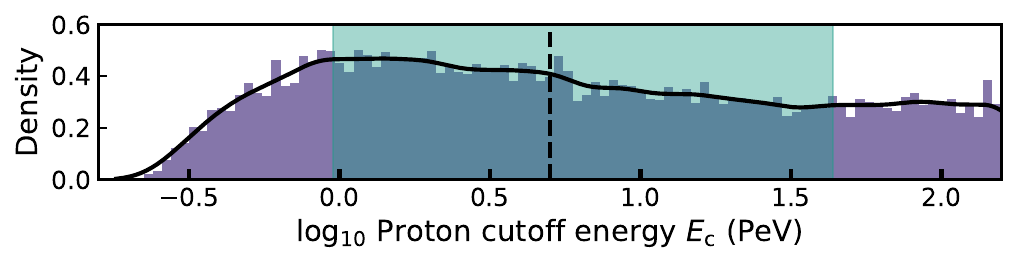}
    \caption[Schematic of proton cutoff energy percentiles]{Posterior distribution of the proton cutoff energy $E_{\mathrm{c}}$. 
The vertical line marks the median (50th percentile, 5~PeV), 
and the light green band shows the 16th–84th percentiles; the high-energy tail is largely unconstrained.}
    \label{fig:ss433:proton_cutoff_energy_schema_percentiles}
\end{figure}

The model predicts a proton energy cut-off around \SI{5}{PeV} for both assumed proton densities, with lower limits of \SI{1}{PeV} (68\% C.L.) and \SI{0.5}{PeV} (95\% C.L.). The posterior flattens at higher energies, making it difficult to establish an upper constraint (see Fig.~\ref{fig:ss433:proton_cutoff_energy_schema_percentiles}).

\section{Multiwavelength Modeling Constraints}
\label{sec:ss433_constraints}

The multiwavelength data place meaningful constraints on the energetics and magnetic field strength of the emission region, but do not tightly constrain the maximum energy of the accelerated electrons in a leptonic scenario. While the observed SED can be reproduced with electron energies well below the PeV scale, electrons at higher energies are not excluded by the data due to rapid radiative cooling and the limited sensitivity of the SED to the high-energy cutoff of the electron distribution.

At magnetic field strengths of $B \sim 10$--$15\,\mu$G, electrons above several tens of TeV cool efficiently via synchrotron radiation, while inverse-Compton scattering becomes increasingly suppressed by Klein--Nishina effects. As a result, the contribution of electrons at the highest energies to the observable photon spectrum is limited. To estimate a physically motivated upper bound on the maximum electron energy, one may apply the Hillas criterion \citep{hillasOriginUltraHighEnergyCosmic1984}, which describes the maximum energy a charged particle can attain while remaining confined within an acceleration region of size $d~pc$ and magnetic field strength $B$:
\begin{equation}
E_{\max} \simeq e Z B d \simeq 10^{16} Z \left(\frac{d}{\mathrm{pc}}\right)\left(\frac{B}{10\mu\mathrm{G}}\right)\,\mathrm{eV}.
\end{equation}
Assuming an acceleration region size of $d \sim \mathrm{pc}$ and a magnetic field strength of $B \sim 10\,\mu$G, this criterion allows for electron energies up to $\sim \times10^{16}\,\mathrm{eV}$. This value should be regarded as a theoretical upper limit, as radiative losses, particularly synchrotron cooling, are expected to reduce the maximum achievable electron energy in practice.

For the leptonic model, the total energy in relativistic electrons required to reproduce the observed SED is $\sim 3\times10^{47}\,\mathrm{erg}$. The emission region is consistent with a magnetic field of $B \sim 10$–$15\,\mu$G. Synchrotron cooling dominates at TeV–PeV electron energies, while inverse Compton, bremsstrahlung, and ionization losses remain subdominant.

For the hadronic model, the total energy in relativistic protons required to reproduce the observed $\gamma$-ray flux depends critically on the local proton density. While the modeling allows for protons to reach PeV energies in the lobes (see Section \ref{sec:mw_sed_modeling_hadronic}), the required proton energy $W_\mathrm{p}$ varies strongly with density: for $n_\mathrm{p} \sim 0.15$–$1.2\,\mathrm{cm^{-3}}$, $W_\mathrm{p}$ spans a factor of 10, with lower-density estimates yielding $W_\mathrm{p} \sim 10^{50}\,\mathrm{erg}$, and higher-density estimates resulting in a significantly smaller energy budget.

In addition to the density estimates from CO and H I observations, an alternative lower-limit estimate can be obtained from the kinetic power of the SS\,433 jets. Using the relation
\begin{equation}
L_{\mathrm{j}} = \frac{1}{2} m_{\mathrm{p}}\, n_{\mathrm{j}}\, v_{\mathrm{j}}^{3}\, A_{\mathrm{j}},
\end{equation}
the proton number density in the jet, \(n_{\mathrm{j}}\) (i.e., the number of protons per unit volume of the jet flow), can be inferred from the observed jet power $L_{\mathrm{j}}$ and jet speed $v_{\mathrm{j}}$, assuming a characteristic cross-sectional area $A_{\mathrm{j}}$ at the emission site, and that the jets consist predominantly of protons.

The true jet cross-section at the location of the extended lobes is not well known, but is likely smaller than the knot size, as seen in X-ray or $\gamma$-ray observations. Assuming a jet half-opening angle $\theta_{\mathrm{open}}$ of $1^\circ$–$2^\circ$, propagating to a distance of $d \sim 38$\,pc (emission region e1; assuming a distance of \SI{5.5}{kpc} to SS 433) from the central source, the jet radius amounts to
\begin{equation}
R_{\mathrm{jet}} \simeq d \, \tan(\theta_{\mathrm{open}}) \sim 0.6-1.3\,\mathrm{pc}.
\end{equation}

This is likely an oversimplified picture: Additional geometric effects arise from precession, with the jets precessing at a half-opening angle $\sim \theta_{\mathrm{prec}} \sim 20^\circ$, which sweeps out a conical surface that could further widen the effective jet channel. On the other hand, re-collimation or confinement by the surrounding W50 cavity could counteract this broadening.

For $L_{\mathrm{j}} \sim 10^{39}\,\mathrm{erg\,s^{-1}}$, $v_{\mathrm{j}} = 0.05c$, and radii of $R \sim 0.6\mathrm{pc} - 1.3\mathrm{pc}$, one obtains
\begin{equation}
n_{\mathrm{j}} \sim 0.007\,\mathrm{cm^{-3}} - 0.033 \,\mathrm{cm^{-3}}.
\end{equation}

The densities derived from this calculation are lower than those inferred from CO and H I measurements of the ambient medium within W50 at the location of the observed $\gamma$-ray emission. However, these estimates depend strongly on the assumed jet properties, scaling as $\sim v_{\mathrm{j}}^{-3}$ with the jet velocity and as $\sim A_{\mathrm{j}}^{-1}$ with the jet cross-sectional area at the emission site.

Taken together, the energetics-based estimate and the gas-density measurements provide complementary constraints: the former determines the particle flux required to sustain the jet power at the emission region, while the latter characterizes the ambient target material available along the jet path. For the densities implied by the energetics-based estimate, the required total proton energy $W_\mathrm{p}$ would be even larger. However, the uncertainties in both the proton density and the age of the system lead to a broad range of possible proton energy budgets, making a purely hadronic origin for the lobe emission energetically challenging, though not conclusively ruled out. These factors highlight the need for further refinement in both density and age estimates.

\section{Discussion and Conclusion}
\label{sec:discussion_conclusion}

Based on the observed $\gamma$-ray morphology and spectrum, the results presented here indicate that the VHE emission from the jet lobes of SS~433 is predominantly leptonic. 
A purely hadronic scenario would require a substantial energy conversion efficiency, with up to $\sim$10\% of the jet kinetic power transferred to relativistic protons. 
While such efficiencies are at the upper end of expectations for astrophysical jets, they are not strictly excluded. 
However, the implied proton energy budget, exceeding $10^{49}\,\mathrm{erg}$, is comparable to or larger than the total kinetic energy injected by the jets over the lifetime of the system, rendering this scenario energetically disfavored.

This conclusion is sensitive to uncertainties in the local target density, the source age, and the total jet power. 
Under more optimistic assumptions for the ambient proton density, the required hadronic energy budget could be reduced. 
The wide range of allowed proton energies reflects the complexity of the system and illustrates that hadronic interpretations remain possible, though energetically challenging given current constraints.

Leptonic models provide a more natural explanation of the observed emission, requiring only sub-percent level energy transfer from the jet to relativistic electrons. 
Inverse Compton scattering of ambient photon fields accounts for the VHE flux with a modest fraction of the jet’s kinetic power. 
Independent constraints on the magnetic field strength from multiwavelength modeling and X-ray synchrotron emission favor values of $B \gtrsim 10\,\mu\mathrm{G}$, consistent with efficient electron acceleration in the extended jet lobes, likely at re-collimation or jet–medium interaction sites.

A far-infrared photon field with a characteristic temperature of $\sim$20~K modifies the inverse Compton component of the broadband SED, primarily affecting the GeV--TeV spectral shape and normalization. 
Direct observational constraints on such a component are limited, as its thermal emission peaks near $\sim$150~$\mu$m, where confusion from diffuse Galactic dust and limited angular resolution dominate. 
As a result, the properties of this photon field remain weakly constrained and introduce an additional source of systematic uncertainty in the inferred electron energy budget. 
Improved $\gamma$-ray measurements, particularly in the GeV band, are expected to provide the most sensitive constraints on its contribution.

The leptonic SED parameters derived in this work are broadly consistent with previous modeling of the SS~433 jet lobes at TeV energies. 
The inferred magnetic field strengths ($B \sim 10$--$20\,\mu$G), electron spectral indices, and lower limits on the electron cutoff energy agree with those reported by the H.E.S.S. collaboration, who favor a leptonic origin for the outer jet emission based on the observed morphology and radiative cooling behavior \citep{h.e.s.s.collaborationAccelerationTransportRelativistic2024}.

At GeV to TeV energies, the one-zone leptonic model presented here is consistent with the emission range in which LHAASO finds inverse Compton scattering by relativistic electrons to be sufficient (up to energies of $\sim$25--30~TeV in their study) \citep{lhaasocollaborationUltrahighEnergyGammarayEmission2024}.
At higher energies above $\sim$100~TeV, LHAASO reports spatially extended $\gamma$-ray emission that is displaced from the jet lobes, suggesting the presence of an additional acceleration site or particle population distinct from the TeV-emitting regions. 
While these observations establish SS~433 as a PeVatron, the limited angular resolution at the highest energies prevents a definitive localization of the ultra-high-energy emission.

The VERITAS observations constrain the leptonic emission associated with the extended jet lobes, while remaining insensitive to the spatially distinct ultra-high-energy component. 
The non-detection of TeV emission from the central region by IACTs places important constraints on inner-jet acceleration scenarios and supports the interpretation that different physical processes dominate particle acceleration at TeV and PeV energies.

In summary, the VHE emission from the extended jet regions of SS~433 is best explained by a predominantly leptonic scenario characterized by efficient electron acceleration and inverse Compton scattering.
While the multiwavelength modeling does not require electrons to be accelerated to PeV energies, such energies cannot be excluded by the present data due to rapid radiative cooling and the limited sensitivity of the SED to the high-energy cutoff of the electron distribution.
At the same time, hadronic models for the jet lobes place constraints on the required proton energy budget, and do not uniquely favor PeV proton acceleration in these regions.
The UHE $\gamma$-ray emission detected closer to the central system likely traces a separate acceleration mechanism, spatially distinct from the TeV-emitting lobes, and may involve a different particle population reaching higher energies. VERITAS observations firmly establish SS~433 as a site of efficient particle acceleration to very-high energies, with steady TeV emission detected from the extended jet lobes in the energy range \SI{0.8}{TeV}--\SI{25}{TeV} and no evidence for emission or variability from the central binary on orbital or precessional timescales. These results highlight the importance of future observations with improved sensitivity and angular resolution, particularly with the Cherenkov Telescope Array Observatory, which will be crucial for resolving the emission morphology in greater detail, testing jet–environment interaction models, and exploring possible connections to high-energy neutrinos in a multi-messenger context.
In principle, hadronic interactions in regions where PeV protons are accelerated could produce neutrinos, which could be probed by current instruments such as IceCube or KM3NeT, providing a direct test of the highest-energy particle acceleration in SS~433.

\newpage

\section{Acknowledgments}
This research is supported by grants from the U.S. Department of Energy Office of Science, the U.S. National Science Foundation and the Smithsonian Institution, by NSERC in Canada, and by the Helmholtz Association in Germany. This research used resources provided by the Open Science Grid, which is supported by the National Science Foundation and the U.S. Department of Energy's Office of Science, and resources of the National Energy Research Scientific Computing Center (NERSC), a U.S. Department of Energy Office of Science User Facility operated under Contract No. DE-AC02-05CH11231. We acknowledge the excellent work of the technical support staff at the Fred Lawrence Whipple Observatory and at the collaborating institutions in the construction and operation of the instrument.

\software{Eventdisplay v491.0 \citep{maierEventdisplayAnalysisReconstruction2017}, V2DL3 v0.6.0 \citep{birdV2DL3VERITASVEGAS2022}, NAIMA v0.10.2 \citep{zabalzaNaimaPythonPackage2016}, Gammapy v1.3 \citep{aceroGammapyPythonToolbox2022}}, Gamera  \citep{hahnGAMERASourceModeling2022}.

%




\appendix

\section{Gammapy FoV Background Method}
\label{sec:gammapy_ana}

The FoV background estimation technique is used as a background estimation method for the VERITAS analysis.
This approach relies on the creation of background models from $\gamma$-ray free data. 
The background models are generated from archival VERITAS observations with quality cuts based on realistic analysis data selection cuts for different observing parameters (azimuth, elevation, $\gamma$-hadron separation cut, epoch, night sky background).
During the background model generation, observations with bright and extended $\gamma$-ray sources are excluded and sources in the VERITAS catalogue are masked with a 0.3 degree radius. 
The background rate is calculated in a (RA, Dec) aligned coordinate system with dependency on energy, and 2D field-of-view coordinate. 

The resulting background models are then attached to the DL3 fits files based on the closest match between observing parameters and background model parameters.

In the analysis, the background is estimated on a run-per-run basis, by fitting the background model outside of the regions of interest and masked regions to closest match the observed counts. Additional energy cuts and maximum offset cuts are applied in this step.

\section{Systematic Bias Estimation Method}  
\label{app:systematic_bias}  

An independent procedure is employed to assess systematic uncertainties and correct for potential biases associated with the 3D maximum-likelihood analysis. Such biases may lead to over- or underestimation of fluxes and significances and can arise from limitations in the background model, including coarse binning in azimuth and zenith, as well as from data quality evaluation procedures.  

To quantify these effects, mimic datasets are generated to reproduce the observational conditions of the analyzed region, including distributions in zenith and azimuth, night-sky background levels, exposure, and instrument configuration. These datasets are derived from $\gamma$-ray-free observations and are processed using the same low-level analysis and DL3 conversion procedures as applied to the real data.  

For each mimic dataset, the background is estimated and compared to the known input, enabling the evaluation of systematic deviations as a function of energy and spatial position. The mimic datasets are stacked to closely match the exposure of the real observations, ensuring that any observed differences reflect analysis-induced biases rather than variations in observational coverage. The background is modeled using a power-law correction across the field of view, and residuals between the predicted background and the observed counts are interpreted as systematic bias.  

The bias is quantified by computing the mean excess across multiple mimic datasets, and corrections are applied to the real dataset to account for these deviations. Significances and flux estimates are subsequently adjusted according to the derived bias and its associated uncertainty. The systematic uncertainty of the analysis method is estimated from the variance of the mimic datasets and is incorporated into the total flux error budget. This approach ensures that the reported fluxes and significances are robust against method-dependent systematic effects and that the uncertainties reflect both statistical and systematic contributions.  

In this way, potential flux biases are evaluated within the eastern and western jet emission regions, as well as in the central region of SS~433. Integration radii of \SI{0.3}{\degree} are employed for this purpose. Fluxes are estimated using a power-law spectral model with slopes taken from the best-fit results, and annular regions surrounding each jet emission zone are used for the estimations.  

The derived flux upper limits range approximately from  
\[
F_{\mathrm{UL}}(95\,\%\:\mathrm{C.L.}) = 1 \times 10^{-13}\,\mathrm{s}^{-1}\,\mathrm{cm}^{-2}  
\]  
to  
\[
F_{\mathrm{UL}}(95\,\%\:\mathrm{C.L.}) = 3 \times 10^{-13}\,\mathrm{s}^{-1}\,\mathrm{cm}^{-2}  
\]  
at locations coinciding with the best-fit emission models. Considering the full extent of the exclusion regions employed in the analysis, the largest flux upper limit is observed in one annular region around w1, with a value of  
\[
F_{\mathrm{UL}}(95\,\%\:\mathrm{C.L.}) = 4.6 \times 10^{-13}\,\mathrm{s}^{-1}\,\mathrm{cm}^{-2}.  
\]  
The mean value across all four regions (e1, e2, w1, w2) is determined to be  
\[
F_{\mathrm{UL}}(95\,\%\:\mathrm{C.L.}) = 3.1 \times 10^{-13}\,\mathrm{s}^{-1}\,\mathrm{cm}^{-2}.  
\]  

These results are consistent with the statistical uncertainties derived from the main analysis. Consequently, no further bias correction is applied, and the total uncertainties are considered to be adequately represented by the statistical errors combined with the standard VERITAS systematic flux uncertainty of $\sim$\SI{25}{\percent}.

\bibliography{references.bib}
\end{document}